\documentclass[journal]{IEEEtran}
\usepackage{listings}
\usepackage{booktabs} 
\usepackage{color}
\usepackage{multirow}
\usepackage{array}
\usepackage[pdftex]{graphicx}
\usepackage{comment}
\graphicspath{{./fig/}}
\usepackage{color}
\usepackage[bookmarks=false]{hyperref}
\usepackage{xcolor}
\usepackage{subfigure}
\usepackage{bm}
\usepackage{enumitem} 
\usepackage{cite}
\usepackage{amsmath,amssymb,amsfonts}
\usepackage{algorithmic}
\usepackage{graphicx}
\usepackage{textcomp}
\usepackage[linesnumbered,ruled,vlined]{algorithm2e}
\usepackage{mathtools}
\usepackage{cases}

\usepackage{tikz}
\newcommand*\circled[1]{\tikz[baseline=(char.base)]{
		\node[shape=circle,draw,inner sep=0pt] (char) {#1};}}

\DeclareGraphicsExtensions{.pdf,.jpeg,.png}
\begin{document}
%
\title{AMF-Placer~{2.0}: Open Source \\ Timing-driven \underline{A}nalytical \underline{M}ixed-size Placer \\ for Large-scale Heterogeneous \underline{F}PGA}

\author{\IEEEauthorblockN{Tingyuan Liang, Gengjie Chen, Jieru Zhao, Sharad Sinha and Wei Zhang
	\thanks{Manuscript received XXXX; revised XXXX; accepted XXXX. Data of publication XXXX; Date of current version XXXX. This work was supported by the Hong Kong Research Grants Council General Research Fund under Grant 16213521.}}

	\IEEEauthorblockA{\textit{\IEEEauthorrefmark{1}ECE Department, Hong Kong University of Science and Technology;}\\ \textit{\IEEEauthorrefmark{2}CSE Department, Chinese University of Hong Kong} \\\textit{\IEEEauthorrefmark{3}CSE Department, Shanghai Jiao Tong University;} \textit{\IEEEauthorrefmark{4}CSE Department, Indian Institute of Technology Goa}\\
		tliang@connect.ust.hk, gjchen@cse.cuhk.edu.hk, zhao-jieru@sjtu.edu.cn, sharad@iitgoa.ac.in, eeweiz@ust.hk}
}
%
%
%

%
%

\markboth{IEEE TRANSACTIONS ON COMPUTER-AIDED DESIGN OF INTEGRATED CIRCUITS AND SYSTEMS, VOL. XX, NO. XX, XXX 2023}%
{Shell \MakeLowercase{\textit{et al.}}: Bare Demo of IEEEtran.cls for IEEE Journals}
%



\maketitle


\begin{abstract}
Modern field-programmable gate arrays (FPGAs) may feature critical path portions of designs prearranged into movable macros during synthesis. These movable macros, with constraints of shape and resources, pose a challenge for mixed-size placement in FPGA designs that previous analytical placers cannot handle. Additionally, general timing-driven placement algorithms face challenges when dealing with real-world application designs and ultrascale FPGA architectures. To address these challenges, we present AMF-Placer~{2.0}, an open-source FPGA placer that supports mixed-size placement of heterogeneous resources. Building on AMF-Placer 1.0, AMF-Placer~{2.0} incorporates new techniques for timing optimization, including an effective regression-based timing model, placement-blockage-aware anchor insertion, TNS/WNS-aware timing-driven quadratic placement, and sector-guided detailed placement. It is evaluated by a set of the latest large open-source benchmarks from various domains for AMD Xilinx Ultrascale FPGAs. Experimental results  indicate that AMF-Placer 2.0 achieves critical path delays that are on average only 2.3\% and 0.69\% higher than those achieved by commercial tool AMD Xilinx Vivado 2020.2 and 2021.2, respectively. Furthermore, the average runtime of the placement procedure in AMF-Placer 2.0 is 7.0\% and 11.5\% lower than that of AMD Xilinx Vivado 2020.2 and 2021.2, respectively. Although limited by the absence of detailed information of devices and designs, AMF-Placer~{2.0} is the first open-source FPGA placer that can handle timing-driven mixed-size placement for practical complex designs with various FPGA resources and achieve comparable quality to the latest commercial tools.
\end{abstract}


\IEEEpeerreviewmaketitle

\section{Introduction}\label{intro}

\begin{figure}[t]
	\setlength{\abovecaptionskip}{0cm}
	\setlength{\belowcaptionskip}{-0.5cm}
	\centering
	\includegraphics[width=\linewidth]{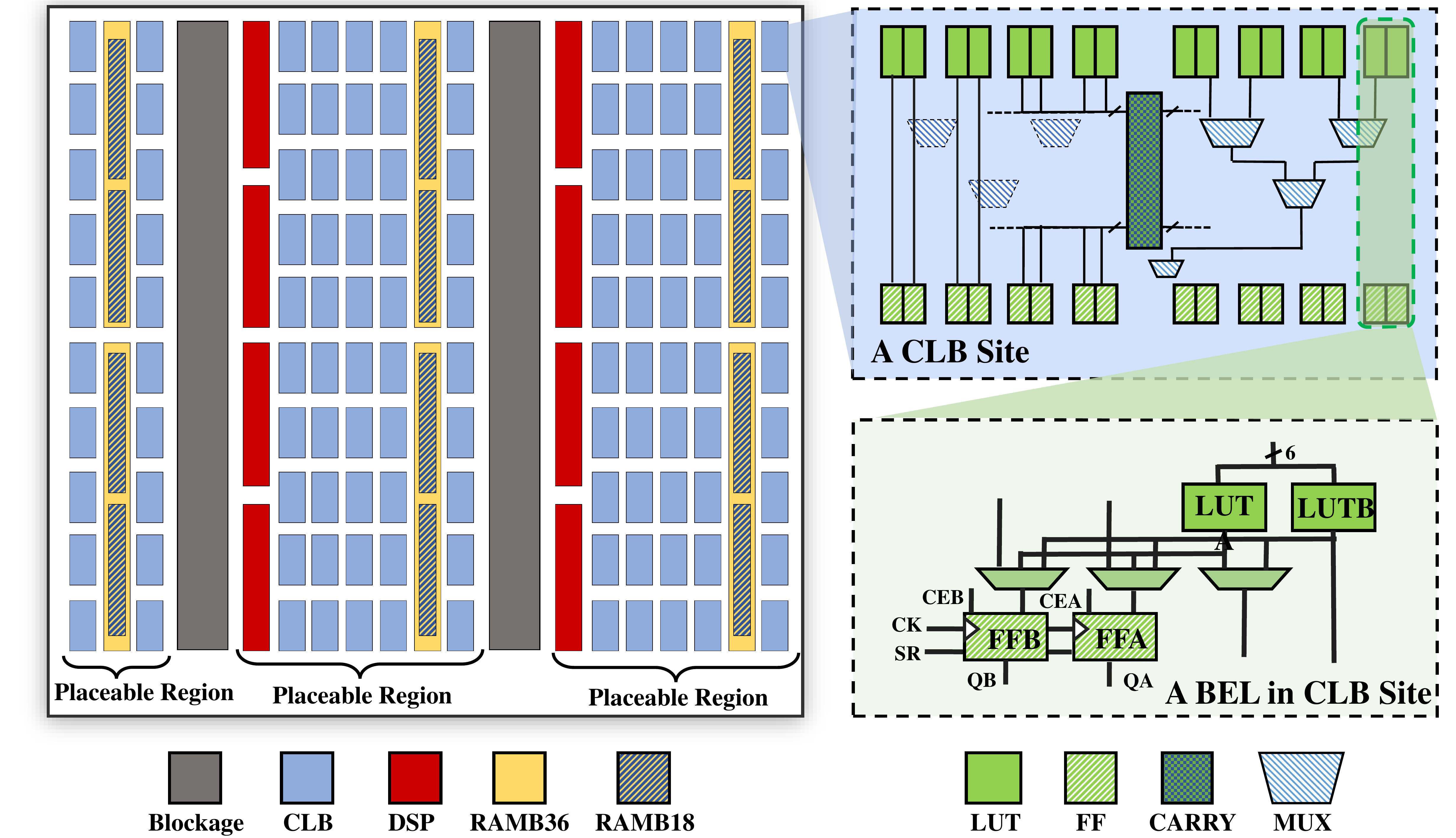}
	\caption{Example of Xilinx Ultrascale FPGA device, a CLB site, and a BEL in it} 
	\vspace{-0.5cm}
	\label{FPGA}
\end{figure}

An Field-Programmable Gate Array (FPGA) is a type of integrated circuit that can be reconfigured by users after manufacturing. The latest island-style FPGAs, such as the columnar and heterogeneous design shown in Fig.\ref{FPGA}, feature a 2D array of configurable sites, with each site consisting of basic elements of logic (BELs)\cite{ultra}. Configurable logic block (CLB) sites, for example, are made up of BELs such as look-up tables (LUTs), flip-flops (FFs), multiplexers (MUXs), and carry chains (CARRYs). Other sites may contain larger, heterogeneous BELs, such as digital signal processors (DSPs) and block random access memories (BRAMs).

During FPGA placement, the netlist generated by logic synthesis should be placed on discrete sites on the FPGA device. The goal is to realize shorter routing wirelength, less congestion regions and better timing, under the constraints of the device architecture. Typically, this placement flow includes the following steps: (1) initial placement/floorplanning to generates a very rough placement; (2) global placement to find optimal locations for the elements to optimize wirelength and timing under the resource constrains; (3) packing and legalization to exactly map each element to a valid FPGA site; (4) detailed placement to resolve the worst cases locally and optimize the metrics. 

With advancement in semiconductor, FPGAs have increased in size as well as the variety of resources available on them and the overall architecture. Meanwhile, FPGA applications become much more complex and denser. These factors bring many new challenges to the placement flow.

\begin{figure}[t]
	\setlength{\abovecaptionskip}{0cm}
	\setlength{\belowcaptionskip}{-0.5cm}
	\centering
	\includegraphics[width=0.95\linewidth]{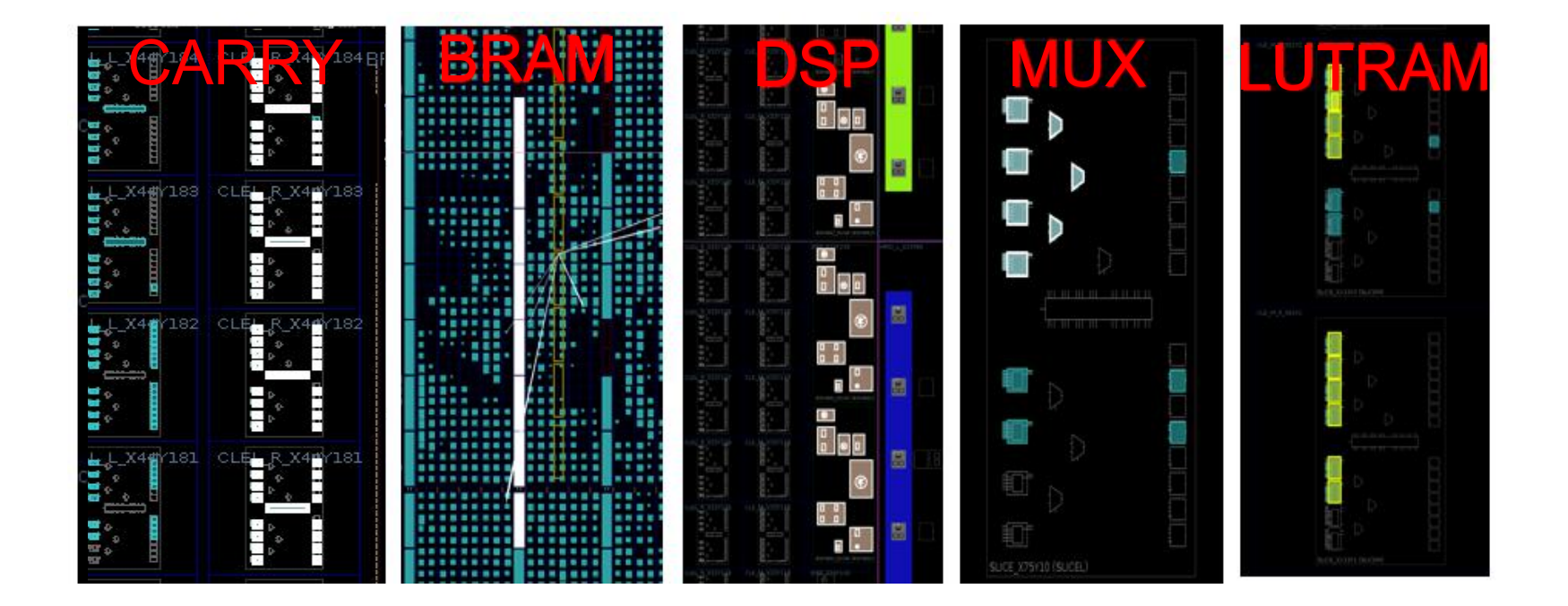}
	\caption{Example of various types of macros with shape constraints: the macros are highlighted. They are combinations of logic blocks, \textbf{occupying multiple BELs}, with constraints of their relative locations.} 
	\vspace{-0.5cm}
	\label{MACROS}
\end{figure}

Due to the upstream optimization, macros with constraints of shape and resource~\cite{whymacro}~\cite{whymacro1} might be generated, like the examples shown in Fig.~\ref{MACROS}. On Ultrascale FPGA architecture~\cite{UG904}, \textbf{standard cell} denotes the smallest, indivisible, representable instance, \textbf{occupying single BEL}, in the design netlist. Meanwhile, \textbf{macro} denotes a fixed group of multiple standard cells, \textbf{occupying multiple BELs}, with constraints of their relative locations. For example: (1) 1 MUX and 2 LUTs connected to it should be treated as a macro; (2) a BRAM, without cascading with other BRAMs, is a standard cell;  (3) 3 cascaded DSPs should be regarded as a macro. On the FPGA device, a macro might require multiple BELs spanning sites. Moreover, macros will lead to high interconnection density. These common scenarios in modern FPGA designs are seldom considered in previous exploration of FPGA placement. 

\subsection{Related Works}\label{relatedWorks}
Some FPGA placers\cite{Vtr}\cite{sa2}, e.g., VTR, are based on simulated annealing (SA), which might lead to long placement runtime when the input netlist is large. Thus, analytical solutions using numerical approaches were proposed to solve the placement with high scalability and quality\cite{anl0}. Gort and Anderson\cite{anl0} presented an analytical FPGA placer HeAP, which demonstrated a 7.4$\times$ runtime advantage with 6\% better placement quality compared to the SA placement algorithm of VPR 5.0. Chen~et~al.~\cite{anl1} proposed analytical placement solution with efficient and effective packing achieves 50\% shorter wirelength, with an 18.30$\times$ overall speedup compared to VPR 7.0. During ISPD 2016/2017 contest, a series of analytical placers, e.g., UTPlaceF\cite{UTPlaceF}\cite{UTPlaceF2} , RippleFPGA\cite{ripplefpga}, GPlace~\cite{GPlace} and NTUfplace~\cite{kuo2017clock}, were inspired with the consideration of congestion and clock constraints and they showed promising performance on the contest benchmarks. Later in 2017, LIQUID\cite{LIQUID} was proposed with analytical solution based on gradient-guided algorithm. ElfPlace~\cite{anl5} cast the placement density cost to the potential energy of an electrostatic system which tried to include various cost metrics in one nonlinear model to be optimized. However, the synthetic ISPD 2016/2017 benchmarks~\cite{ispd2016}~\cite{yang2017clock} have some limitations. For example, the randomly-generated netlists contain impractical interconnections. Each 36Kb Block-RAM has only 6 input nets and 1 output net, and many registers are unnecessarily duplicated without consideration of timing and fanout. Furthermore, the netlists have no design hierarchy, which results in a relatively even distribution of timing criticality. Additionally, the benchmark lacks widely-used instances like CARRY, MUX, LUTRAM, and 18Kb Block-RAM. These limitations make the benchmark less representative of real-world scenarios.

Timing-driven FPGA placement is critical for achieving optimal timing quality in FPGA design implementation. However, this type of placement is challenging due to the complex interaction among routing wirelength, routability, and timing quality. Improper emphasis or neglect of the reduction of estimated wirelength, such as half perimeter wire length (HPWL), can result in routing failure, longer routing wirelength, and worse timing~\cite{liao2022dreamplace}\cite{hyun2019accurate}. Additionally, the discrete heterogeneous resource constraints of FPGAs pose significant challenges that are not accounted for in existing timing-driven solutions for ASIC placement\cite{asic2010,asic2015,2016asicTodas,asic2017,ajayi2019toward,asic2020,asic2021,asic2022, liao2022dreamplace}. To address these challenges, researchers have proposed various solutions.
Chen and Chang~\cite{chen2015timing} proposed a local-routing-architecture-aware timing cost function in the analytical FPGA placement problem.
Dhar,~et~al.~\cite{dhar2017effective} introduced an effective detailed placement based on the shortest path algorithm, which has been adopted by many solutions.
Lin,~et~al.~\cite{lin2020analytical,lin2021timing} proposed an efficient delay model and timing-driven placement that consider clock constraints.
Nikoli\'{c},~et~al.~\cite{fpga2022} developed an efficient ILP-based detailed placer that moves a carefully selected subset of LUTs to improve timing.

Real-world applications are becoming increasingly complex, and FPGA devices used to handle them have many architecture constraints. These constraints include placement blockages, packing legalization, clock legalization, and fixed macro shapes.
As a result, timing-driven placement is becoming more difficult. Advanced algorithms and tools are needed to optimize placement while still meeting these complex restrictions. The existing challenges will be discussed in Section~\ref{Motivation}.

\subsection{Motivation}\label{Motivation}
%
%
%
%

\subsubsection{Challenges of FPGA Timing-driven Placement}\label{challenges1}
In addition to the challenges we discussed in the introduction of AMF-Placer 1.0 \cite{liang2021amf} regarding mixed-size placement, there are still many unresolved challenges in timing-driven FPGA placement. These challenges span a wide range of perspectives and have not been addressed by previous works:

\begin{itemize}
	\item Global placement algorithms in previous works~\cite{chen2015timing,lin2020analytical,lin2021timing} are mainly guided by TNS (total negative slacks), which is a scalar value and cannot capture the negative slack distribution among the nets. Neglecting the impact of WNS (worst negative slack) during global placement might result in suboptimal WNS results. 
	\item The latest detailed placement algorithms are commonly based on the shortest path algorithm~\cite{dhar2017effective}, which suffers from low-efficiency identification of candidate locations for involved instances.
	\item The complexity of application netlist and FPGA architectures has been raised dramatically. Complex FPGA architecture factors (e.g., placement blockage in Fig.~\ref{FPGA}) and mixed-size designs cannot be handled properly by the existing solutions~\cite{chen2015timing,lin2020analytical,lin2021timing}.
\end{itemize}

\subsubsection{Impact of Macro Instances on Timing Optimization }\label{challenges2}

Mixed-size instances pose a significant challenge for timing optimization due to several factors.

\begin{itemize}
\item Large macros can cause disruptions in the placement of other instances in critical timing paths, even if the macros themselves are not part of those paths. This is because macros often require multiple sites or BELs. For instance, a CARRY macro can occupy over 128 BELs (including LUTs, FFs, and CARRYs) and 8 CLB sites, leading to resource conflicts with many other instances.
\item Macros, unlike standard cells, typically have a larger number of pins and nets connected to other instances. For example, the CARRY macro mentioned earlier may have over 300 nets connected to it outside of the macro. This is due to its inclusion of 64 LUT6 cells, each with 6 fanin pins. This high fanin and fanout of macros can lead to a significant number of intersections in critical timing paths, making timing optimization challenging. Moving an instance to optimize the timing of one path may result in a significant degradation of timing in many other paths.
\end{itemize}

\subsection{Contributions}\label{contributions}

AMF-Placer~{1.0}~\cite{liang2021amf} enables efficient mixed-size FPGA placement with parallelized techniques including: (1) simulated-annealing-based floorplanning; (2) quadratic placement with interconnection-density-aware pseudo nets and legalization-oriented anchors; (3) cell spreading algorithm with utilization-guided search window and deadlock-free area supply control; (4) and progressive macro legalization. AMF-Placer 2.0 takes into account the practical demands of timing quality and the complexities of real-world applications with hierarchies that involve elements with shape constraints. It builds on the foundation laid by AMF-Placer 1.0 and offers several new essential features, including:

\begin{itemize}
	\item a set of timing optimization algorithms that do not require static timing analysis (STA), such as path-length-aware SA-based floorplanning and parallelized CLB packing with timing factors considered.
	\item an efficient piecewise regression model of pin-to-pin delay, utilized by our integrated light-weight parallelized timing analysis engine.
	\item a placement-blockage-aware optimization scheme that identifies the potential negative interference of placement blockage with long paths. It spreads instances in specific regions and inserts placement anchors for the instances in the target paths to reduce cross-blockage routing.
	\item a WNS/TNS-aware timing-driven global placement algorithm based on quadratic programming and proper exploitation of pseudo-nets with slack-guided weights. This algorithm achieves multi-objective optimization of WNS, TNS, and wirelength.
	\item  a sector-guided detailed placement algorithm that can efficiently identify instance movements with promising timing benefits.
\end{itemize}

The source code and Wiki of our proposed AMF-Placer~{2.0} and involved open-source benchmarks are available at \href{https://github.com/zslwyuan/AMF-Placer}{https://github.com/zslwyuan/AMF-Placer}.

\section{Preliminaries}

In this section, we describe the mixed-size placement problem in FPGA scenarios and our analytical placement framework.

\subsection{Characteristics of Mixed-size Design and Ultrascale Device Architecture} \label{macrofeature}

In Fig.\ref{MACROS}, we can see that standard cells in a macro must be placed in adjacent sites in the same column to meet downstream flow requirements. Typically, each macro includes one type of core cell, such as CARRY cells, MUX cells, LUTRAM cells, DSP cells, or BRAM cells. In addition to core cells, macros may also include peripheral LUT/FF cells directly connected to the core cells. Macro types can be primarily classified into five categories, each with distinct characteristics:
\begin{itemize} 
	\item The CARRYs connected with carry in/out port should be extracted as a macro, along with the LUTs and FFs directly connected to them. Furthermore, to enable the routing of some input pins of CARRY, which connect to signals outside the CLB site, some corresponding LUT slots in the same CLB site should be occupied by non-logic route-thru LUTs, which are not in the original netlist. Similarly, FF slots in CLBs may be unavailable due to routing resource contention in CARRY macros.
	\item A MUX with its two input standard cells, which could be two LUTs or two other MUXes, should be extracted as a macro. MUX macros may lead to route-thru usage of LUTs or disable external interconnection of some FFs due to the routing of selection signals.
	\item LUTRAM standard cells, which share input net for read/write address and data bits, should be extracted as a macro. LUTRAM macros should be located in SLICEM columns of the device. 
	\item For DSPs and RAMs, they might be cascaded to handle larger demand for computation and storage. The standard cells in one of these macros are interconnected by the nets of their cascaded input/output signals.  
\end{itemize}

AMF-Placer~{2.0} inherits the ability of AMF-Placer~{1.0}~\cite{liang2021amf} to detect the macros in the design netlist and generates placeholders to occupy resources and meet internal routing constraints. There are some other minor macros defined by vendor primitives~\cite{whymacro1}, which are out of the scope of this work. More details are available in the device documentations\cite{clb,bram,dsp}.

In modern FPGA devices, placement blockages separate the device into several available placement regions, which may be introduced by IO banks (e.g., GPIOs and PCIe interfaces) or die boundaries, as shown in Fig.\ref{FPGA}.  The delays of nets across the blockage region can be relatively higher than the delays of the nets routed within the general placement region. This problem is critical in timing-driven placement but is often ignored in existing works that focus on wirelength-driven placement. For example, the placement blockages of PCIe interfaces are omitted entirely in the device information of the benchmark ISPD 2016/2017\cite{ispd2016}~\cite{yang2017clock}.

\subsection{Problem Formulation}

The placement of the instances in a FPGA-based design can be formulated as a hypergraph $H = (V, E)$ placement problem. Let vertices $V = \{v_1 , v_2, \dots , v_n \}$ represent $n$ instances in the design netlist and hyperedges $E =\{e_1, e_2, \dots , e_m\}$ represent $m$ nets. Let $x_i$ and $y_i$ be the $x$ and $y$ coordinates of the center of the instance $v_i$ during placement, respectively. As mentioned in Section \ref{intro}, the instances can be categorized into two types, i.e., standard cells and macros, and both of these two types could be movable or fixed according to the design constraints.  One of the most common objective functions for placement is the
sum of half-perimeter wirelength (HPWL) over all nets, i.e., the defined $E$. By properly inserting weighted pin-to-pin pseudo nets, AMF-Placer~{2.0} can integrate the timing objective into the conventional wirelength-driven placer. The mixed-size FPGA placer is responsible for determining the position of each movable instance (i.e., $x_i$ and $y_i$) to minimize the objective function, which comprises wirelength and timing terms, while adhering to technology and region constraints.

\subsection{The Framework of AMF-Placer~{2.0}}
\begin{figure}[t]
	\setlength{\abovecaptionskip}{-0.0cm}
	\setlength{\belowcaptionskip}{-0.5cm}
	\centering
	\includegraphics[width=\linewidth]{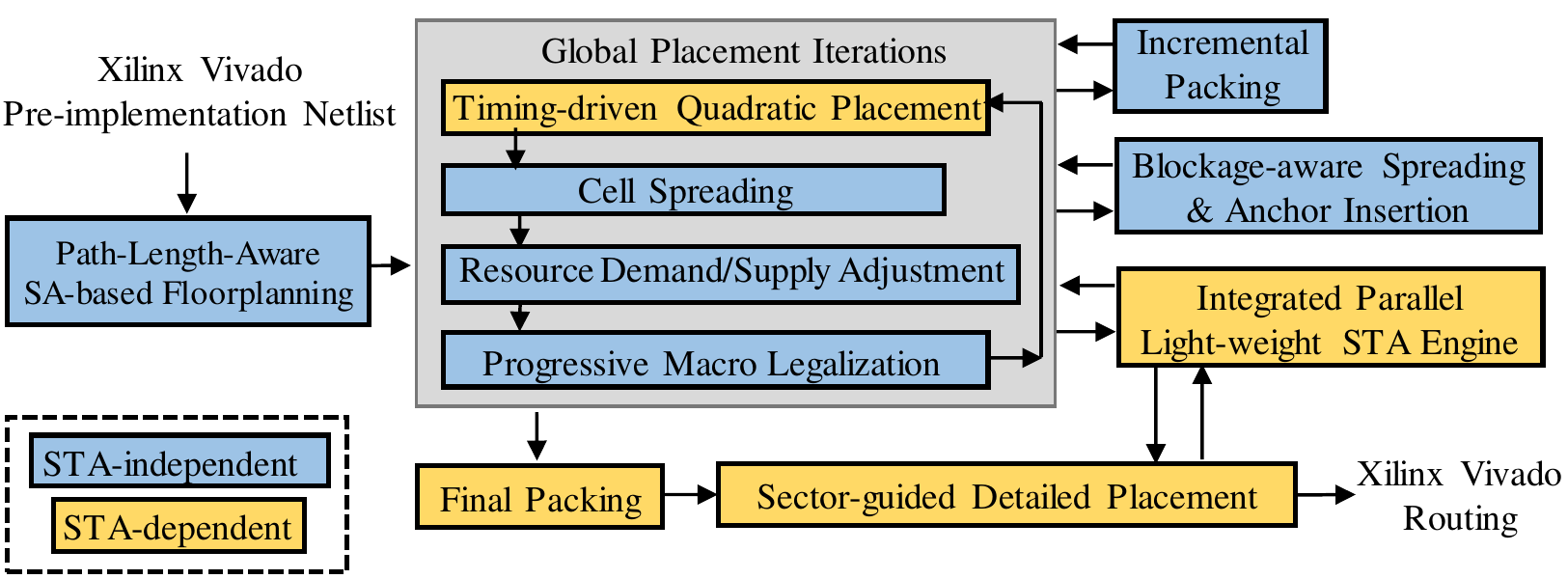}
	\caption{The workflow of AMF-Placer~{2.0}  consisting STA-independent phases and STA-dependent phases } 
	\vspace{-0.5cm}
	\label{outline}
\end{figure}

AMF-Placer~{2.0} is a comprehensive FPGA placement framework as shown in Fig.~\ref{outline}. The input of AMF-Placer is the pre-implementation netlist extracted from AMD Xilinx Vivado. The outputs are the location of each instance on the specific device and a Tcl script for Vivado to consume the generated placement. The proposed placement flow consists of STA-independent phases and STA-dependent phases. 

\subsubsection{STA-independent Phases} 
\quad

During the early stage of placement, instances can be moved extensively, making it challenging to obtain a reliable timing evaluation of the placement. As a result, STA-independent phases focus on early-stage timing optimization to minimize HPWL and pave the way for timing-driven phases. These STA-independent phases include:

{\emph{(1.1) Initial Floorplanning:}}  AMF-Placer~{2.0} begins with simulated-annealing(SA)-based floorplanning of the instance partitions obtained by the  path-length-aware clustering and  connectivity-based partitioning. 

{\emph{(1.2) Blockage-aware Spreading and Anchor Insertion:}} This phase analyzes the connectivity, the timing criticality of the circuits, and the location distribution of the critical paths on the device. It clusters some instances and inserts anchors for these instances to reduce the long cross-placement-blockage routing. To enable the coarse-grained movements of the clusters, instances in specific regions will be spread. 

{\emph{(1.3) Cell Spreading:}} Based on the area demand of instances and the area supply of the devices, instances will be spread from the regions where the demand for resources is outrunning supply, to other regions. We adopt the cell spreading algorithm of  AMF-Placer~{1.0}~\cite{liang2021amf} for mixed-size instances.

{\emph{(1.4) Resource Demand/Supply Adjustment:}} Based on the packing feasibility and routing congestion level, the area demands of instances and the area supply of some regions will be adjusted, to improve the placement quality. This phase is adopted from extended UTPlaceF\cite{parallelpack}.

{\emph{(1.5) Progressive Macro Legalization:}} Mainly adopted from AMF-Placer~{1.0}~\cite{liang2021amf}, each macro will be mapped to multiple potential locations or one exact location according to the confidence. 

{\emph{(1.6) Incremental Packing:}} Adopted from RippleFPGA\cite{ripplefpga}, some LUTs and FFs will be paired as macros to improve the placement quality by identifying CLB internal nets. 

\subsubsection{STA-dependent Phases} \quad

When the wirelength tends to be stable, timing-driven phases will further optimize the timing quality of the placement based on static timing analysis. These timing-driven phases include:

{\emph{(2.1) Timing Model and Timing Analysis:}} Based on the dataset of timing delays, we use a piece-wise function based on non-integer polynomials to fit the distribution of timing delays. Based on this regression model of timing and the ideas of OpenTimer~\cite{huang2015opentimer}, a light-weight parallel static timing analysis engine is implemented.

{\emph{(2.2) Timing-driven Quadratic Placement:}} To determine the next location of each instance, a quadratic optimization problem involving wirelength and timing is solved. In mixed-size placement, interconnection density and legalization are taken into account during quadratic placement. Additionally, timing-oriented pseudo nets are inserted between pins to optimize timing. The strength of these nets is determined by both the local timing slack of related paths and global timing quality.

{\emph{(2.3) Global Packing:}} After the global placement iterations, the next phase involves exact legalization, which assigns each instance to specific sites with a fixed number and type of resources. For instance, BELs, including LUTs, FFs, MUXs, and CARRYs, are mapped to configurable logic blocks (CLBs). This phase focuses on maximizing the internal interconnection of CLBs while taking into account timing factors, under the constraints of available resources and clocking.

{\emph{(2.4) Detailed Placement:}} Global placement and global packing evaluate placement quality from a global perspective, whereas detailed placement focuses on local critical paths that play a smaller role in global objectives. AMF-Placer 2.0 follows a similar placement flow to UTPlaceF \cite{UTPlaceF}\cite{UTPlaceF2},  RippleFPGA \cite{ripplefpga}, and GPlace\cite{GPlace} by performing detailed placement after packing. This work flow can ensure legalization and identify available slots for instance replacement during detailed placement.

Most of the related algorithms for these phases are parallelized. Detailed methodologies  will be illustrated in the following section~\ref{WLimplementation} and~\ref{Timingimplementation}.

\section{Implementation of STA-independent Phases}\label{WLimplementation}

In this section, we will illustrate the implementation of STA-independent phases which do not rely on static timing analysis (STA). Some STA-independent phases are based on AMF-Placer~{1.0}, and we will focus on the illustration of the additional important timing-oriented modifications.

\subsection{Initial Floorplanning}\label{floorplanning}

Large-scale real-world FPGA designs, as shown in Section~\ref{evaluation}, have specific architectures and hierarchies that differ from randomly generated FPGA netlists\cite{ispd2016}\cite{gnl} or small designs\cite{anl0}. Additionally, FPGA resources are divided into discrete regions, and the supply of resources for each type is not uniform across the device. These design and device factors make the initial floorplan critical for later timing optimization, especially for large macros with high fan-in/fan-out.

Previous partitioning/clustering algorithms, such as those in \cite{ripplefpga}\cite{liang2021amf}\cite{patoh}\cite{clockcluster}, do not account for timing factors during partitioning. Assigning weights to nets based on timing slack is a potential approach, but it may be impractical due to unreliable timing slack evaluation at the beginning of placement and the focus on minimizing total negative slack rather than worst negative slack. Besides, weighted hypergraph partitioning algorithms require additional operations, such as computing the sum of weights and sorting edges, which can increase complexity. Additionally, balancing weights across partitions further complicates the optimization problem.

To improve the timing optimization of AMF-Placer, version 2.0 introduces a novel clustering approach based on timing path length. By using breadth-first search (BFS) from the start and end points of timing paths, AMF-Placer 2.0 can easily obtain the distance of each instance to the farthest start or end point, resulting in the maximum length of the paths including that instance.

Next, AMF-Placer 2.0 sorts the instances according to their maximum path length (primary key) and the distance to the farthest start point (secondary key), generating a sorted list of instances. To focus on the instances in the longest paths, a path length threshold is introduced to select only the top 5\% instances in the list. 

\begin{figure}[t]
	\setlength{\abovecaptionskip}{-0.0cm}
	\setlength{\belowcaptionskip}{-0.5cm}
	\centering
	\includegraphics[width=\linewidth]{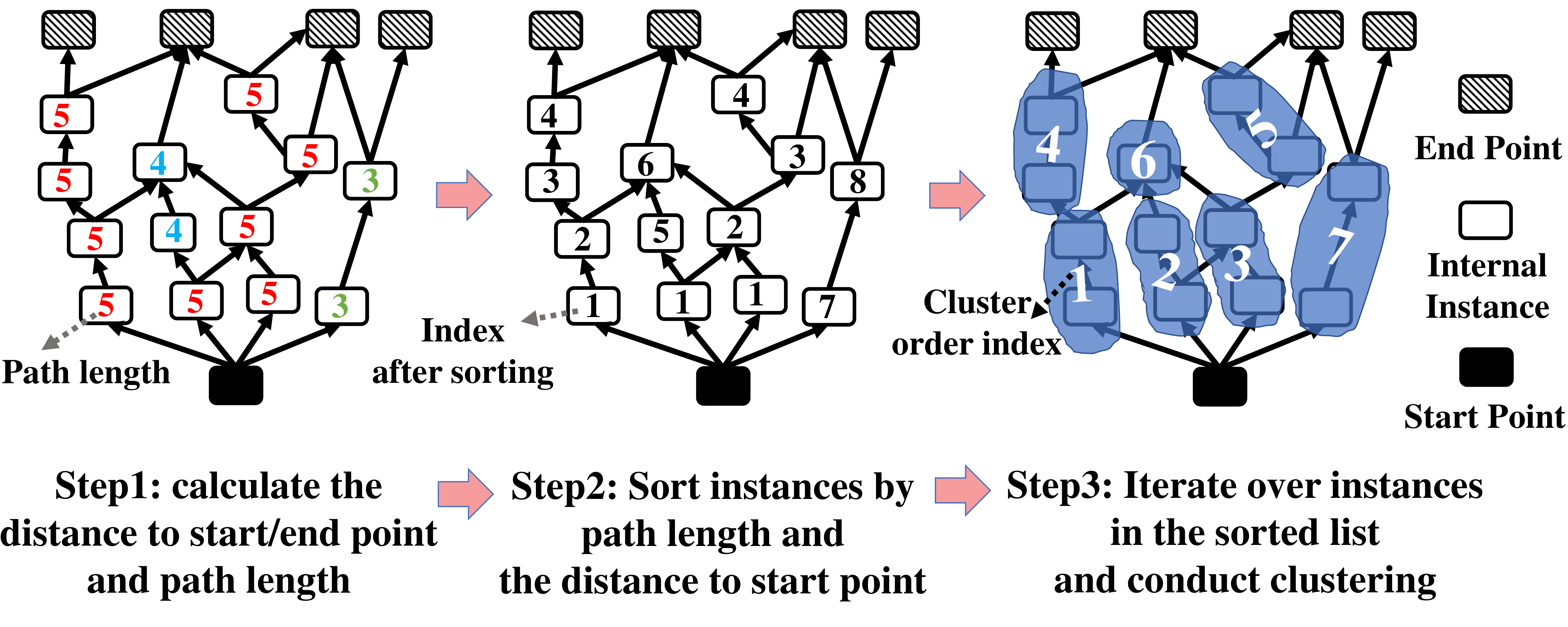}
	\caption{Three steps of clustering guided by path length} 
	\label{initialF}
	\vspace{-0.5cm}
\end{figure}

AMF-Placer 2.0 iterates over instances in the sorted list, clustering each unclustered instance with its unclustered direct fanout instances as shown in Fig~\ref{initialF}. The resulting fine-grained clusters are treated as entities during partitioning, reducing cross-cluster interconnections in critical paths.

AMF-Placer 2.0 does not directly cluster long paths. This is because an instance may be included in multiple paths with similar lengths, and clustering one path could negatively impact the timing of the other paths. Additionally, extremely long paths can still span a wide range in the final placement, and cluster them will limit later optimization. Instead, AMF-Placer 2.0 clusters instances with their direct fanout, realizing a more balanced and effective partitioning solution.

After the generation of fined-grained clusters, the conventional connectivity-based partitioning~\cite{patoh} will be involved to obtain tens of coarse-grained partitions of the design netlist. In this partitioning procedure, each of the fine-grained clusters are treated as an entity node. Finally, a simulated-annealing(SA)-based floorplanning in AMF-Placer 1.0~\cite{liang2021amf} will determine the location of partitions on the FPGA device.

\subsection{Blockage-aware Spreading and Anchor Insertion}\label{blockageSolution}

\begin{figure}[!b]
		\vspace{-0.5cm}
	\setlength{\abovecaptionskip}{0.cm}
	\setlength{\belowcaptionskip}{-0.5cm}
	\centering
	\subfigure[Pure analytical solutions: cell spreading will spread instance from placement blockage and make the solution less effective.]{
		\begin{minipage}[b]{0.45\linewidth}
			\includegraphics[width=\linewidth]{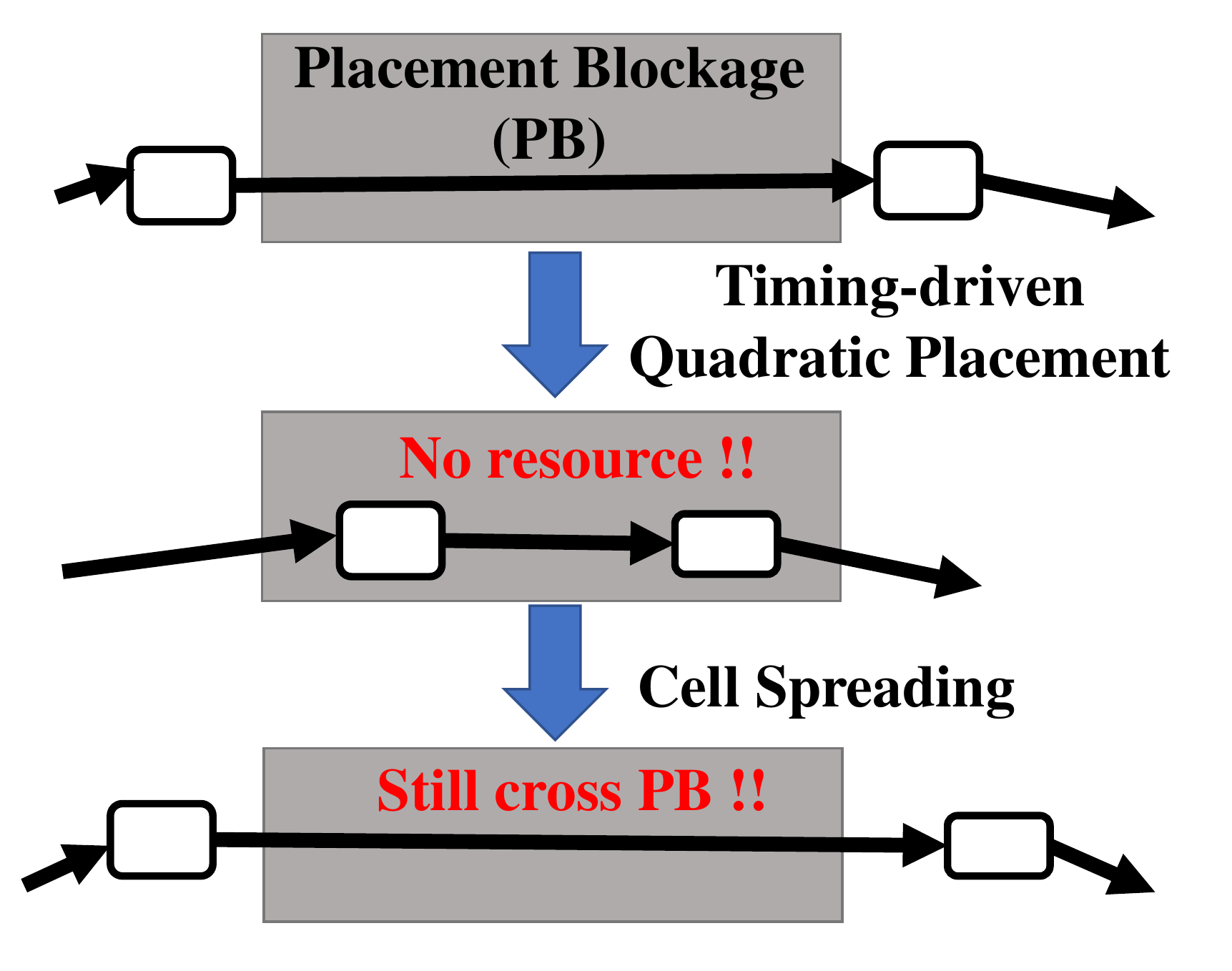} 
			\label{wlNotWork}\vspace{-0.5cm}
		\end{minipage}
	}
	\subfigure[Fine-grained techniques:  moving individual instances cannot effectively improve the critical path delay due to many long high-fanout paths.]{
		\begin{minipage}[b]{0.45\linewidth}
			\includegraphics[width=\linewidth]{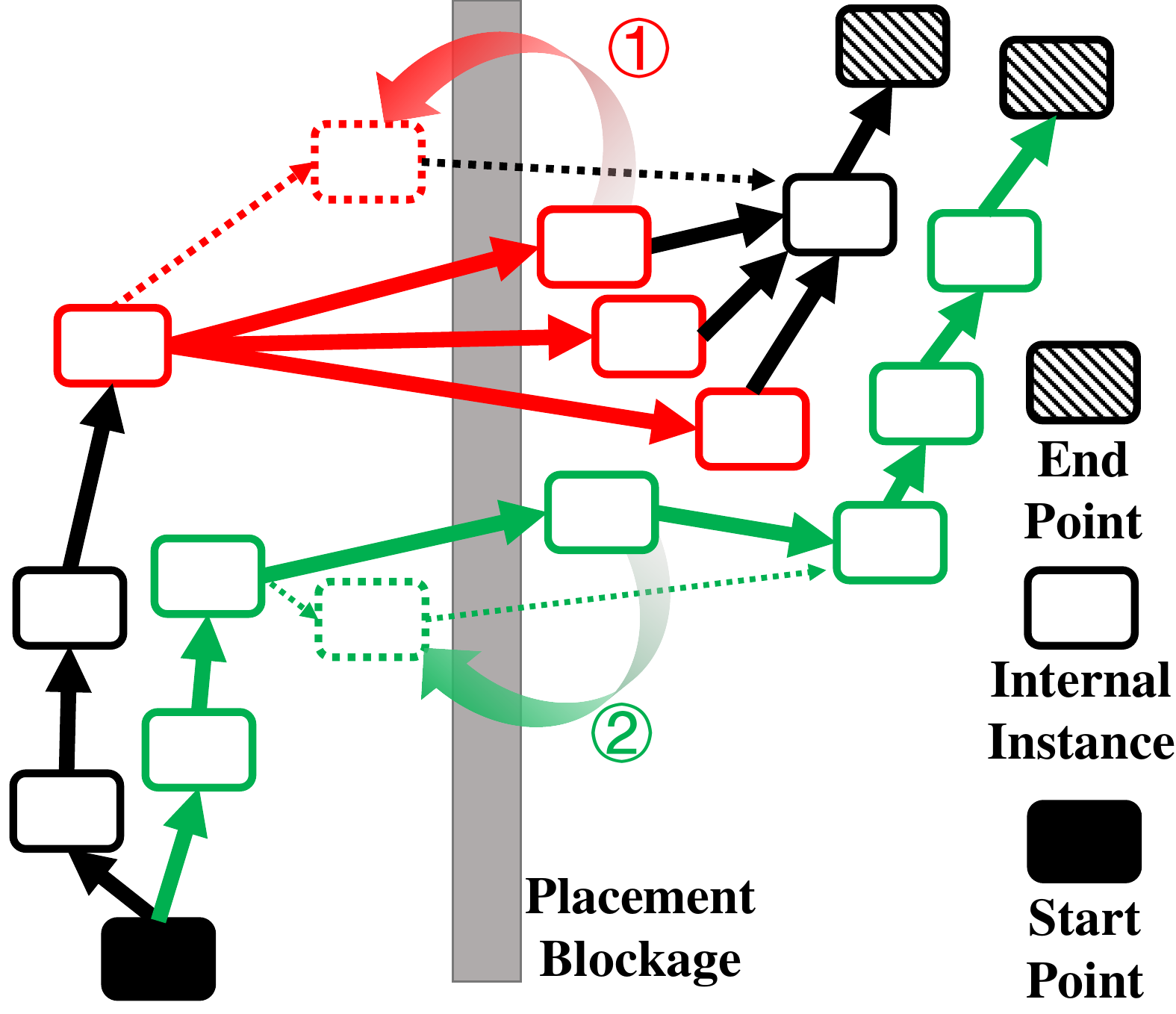} 
			\label{localNotWork}
		\end{minipage}
	}
	
	\caption{Drawbacks of the other timing optimization solutions for the columnar blockage regions} 
	\label{motivationBlk}
\end{figure}

Fig.\ref{wlNotWork} demonstrates that simple timing-slack-based net weights cannot solve the timing problem caused by nets across a blockage region. This is because the reduced distance between instances across the blockage may be reverted by cell spreading or final packing. Furthermore, fine-grained local placement optimization is not effective in resolving this problem due to two main reasons. Firstly, one net spanning blockages may drive multiple sink instances, so resolving the problem for one sink instance may not address the cross-blockage problem for other sink instances, as shown by \circled{1} in Fig.\ref{localNotWork}. Secondly, one net may belong to a long path. While resolving the problem of a pair of instances on the path, the cross-blockage problem for other instances in the path may arise, as shown by \circled{2} in Fig.~\ref{localNotWork}.

To address blockage-related problems during global placement, we propose a coarse-grained solution involving three steps as shown in Fig.~\ref{outline}.

First, critical path circuits are clustered based on a sorted list of instances with maximum path length greater than a threshold. Each instance is contained in only one of the resulting clusters, which are obtained by traversing unclustered direct/indirect successors of target instances in a depth-first search. The maximum size of each cluster is limited to an empirical number, which is set to 20000 in AMF-Placer 2.0.

Second, a target available placement region is selected for each cluster. If the proportion of instances belonging to a cluster in a specific available placement region is greater than 50\%, that region is selected as the target region. Otherwise, the cluster is not assigned to any available placement region and instances in it will be released for the other clusters.

\begin{figure}[t]
	\setlength{\abovecaptionskip}{-0.0cm}
	\setlength{\belowcaptionskip}{-0.5cm}
	\centering
	\includegraphics[width=0.9\linewidth]{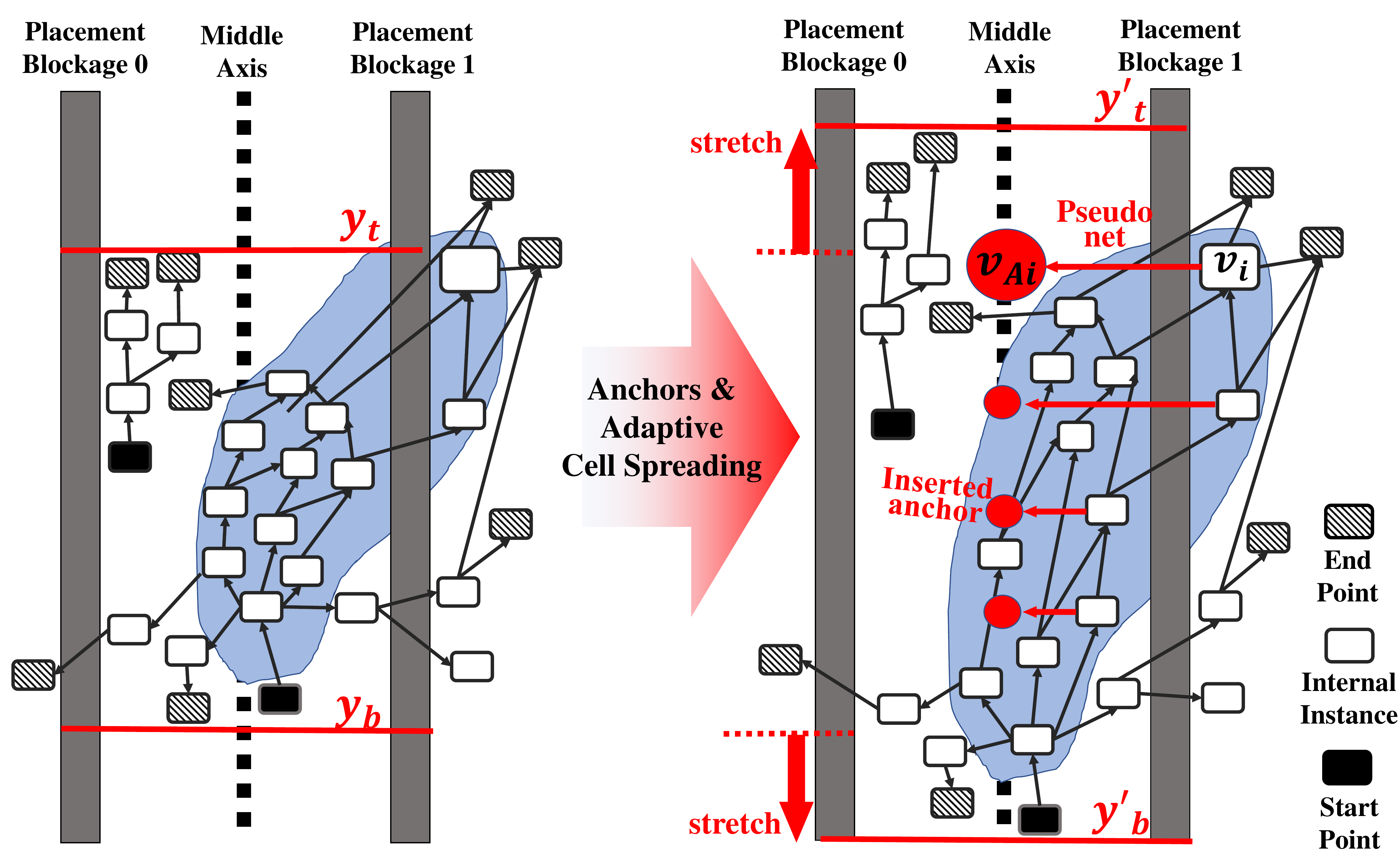}
	\caption{Examples of Blockage-aware Spreading and Anchor Insertion: some of the pseudo nets are omitted for better visualization.}  
	\vspace{-0.5cm}
	\label{blockAwareSol}
\end{figure}

The third step involves guiding the movement of instances in a cluster to its corresponding target region during later placement iterations. In Ultrascale FPGA devices, available placement regions and blockages are typically columnar and horizontally aligned. Accordingly, each instance in the cluster is connected to a corresponding anchor at the horizontal center of the target region. The anchor has the same vertical coordinate as the instance.  An example is shown in Fig.~\ref{blockAwareSol}, where $v_{Ai}$ is an anchor and  $v_i$ is an instance in the cluster. For the timing-driven quadratic placement which will be illustrated in Section~\ref{timingQP}, the weight of the pseudo net $e_b(v_i)$ connecting $v_{Ai}$ and $v_i$ can be formulated as:
\begin{eqnarray}
	w_{e_b}(v_i) =&\beta \times |(x_i-x_{Ai})| \times \text{pinNum}(v_i) \label{guidenet}
\end{eqnarray}
where $\beta$ is a constant hyperparameter and $\text{pinNum}(v_i)$ is the number of the pins of $v_i$. For all the instances assigned to specific target regions, their blockage-aware pseudo nets will be recorded in a set $E_b= \left\{e_b(v_i)\right\}$. According to Eqn.(\ref{guidenet}), even for instances $v_i$ in their target region, pseudo nets with lower weights will still be attached, and when $v_i$ is far from $v_{Ai}$, the corresponding net will be strengthened.

An available target placement region could become highly utilized or congested during placement, so we need to stretch the placement in the target region. This is necessary to ensure that incoming clusters of instances have ample room for placement without causing serious congestion problems.

The formula for calculating the stretch ratio is: $\Delta r_{stretch}=N_{outside}/N_{inside}$, where $N_{outside}$ is the number of instances guided to the target region but not currently in it, and $N_{inside}$ is the total number of instances in the target region  that are not to be guided to other regions.

Let $y_{t}$ be the vertical coordinate of the top instance $v_t$ in the target region and $y_{b}$ be the vertical coordinate of the bottom one. The linearly transformed vertical coordinates for $v_t$ and $v_b$ are calculated as follows:
\begin{eqnarray}
	y_{t}' =& y_{t} + \Delta r_{stretch} \times (y_{t}-y_{b}) / 2  \\ 
	y_{b}' =&y_{b} - \Delta r_{stretch} \times (y_{t}-y_{b}) / 2
\end{eqnarray}
Here, $y_{t}'$ and $y_{b}'$ are the new vertical coordinates for $v_t$ and $v_b$, respectively.

Instances within the target region that have original vertical coordinates between $y_{t}$ and $y_{b}$ will be mapped to new locations using a linear transformation of coordinates, similar to the method shown in Fig.~\ref{blockAwareSol}. In the case that $y_{t}'$ or $y_{b}'$ falls outside the boundary of the FPGA device, slight adjustments will ensure that all instances remain within the device boundary.

The insertion of anchors and specific instance spreading helps to gradually consume the clusters with long paths across blockages by mapping their instances to their corresponding placement regions.
This approach ensures that available placement space is efficiently utilized, while also avoiding congestion problems.

\section{Implementation of STA-dependent Phases}\label{Timingimplementation}

\subsection{Regression-based Timing Model and Timing Analysis}\label{timingModel}

An accurate timing delay model is crucial in timing-driven placement. Some previous solutions for timing delay modeling~\cite{lin2021timing}~\cite{maidee2019open}~\cite{10.1145/3491236} in FPGA placement are based on timing look-up table or ideal routing path. They may suffer from limitations in applicability and accuracy. Some solutions~\cite{lin2021timing}~\cite{martin2019flat}  lack availability since they rely on industry-provided dataset. 

\begin{figure}[t]
	\setlength{\abovecaptionskip}{-0cm}
	\setlength{\belowcaptionskip}{-0cm}
	\centering
	\includegraphics[width=\linewidth]{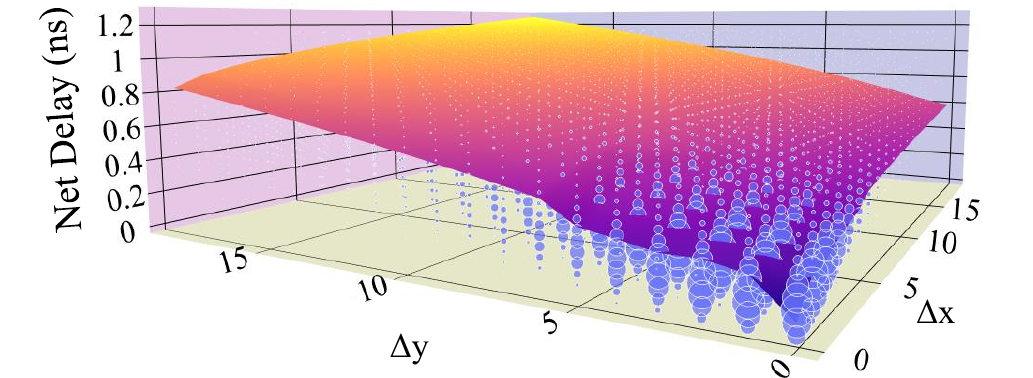}
	\caption{Sample Distribution and Regression Model Surface: The markers of varying sizes depict the density of the samples in our dataset. The surface is generated by the piece-wise function $T_{net}$ with different $\Delta x$ and $\Delta y$ } 
	\label{timingDistribution}
		\vspace{-0.4cm}
\end{figure}

AMF-Placer~{2.0} employs both a regression model and a look-up table. The fixed logic delay ($T_{logic}$) is recorded in the look-up table, while the regression model estimates the variable net delay ($T_{net}$). To create the regression model, we generate a dataset of instance-to-instance timing delays of nets in real-world benchmarks, as described in Section~\ref{experimentSetting}. We randomly extract this data using the Vivado Tcl command "get\_net\_delays" and visualize it in Fig.\ref{timingDistribution}. The dataset consists of $N_{sample}$ samples that represent the locations of instances and the routing delay between them, as determined by the actual routing. In AMF-Placer{2.0}, we set $N_{sample}$ to 10000, which is sufficient for the factors in the regression model to converge. We use a non-integer polynomial function to fit the distribution of timing delays, accounting for the differences in X-Y coordinates of interconnected instances on the FPGA. The formula of the net delay estimation function $T_{net}(e_{i,j})$ is:
\begin{equation}\label{timingRegression}
	\begin{aligned}
	T_{net}(e_{i,j})=&a_0 \Delta x^{b_0}+ a_1 \Delta x^{b_1} 	+ a_2 \Delta y^{b_0}+ a_3 \Delta y^{b_1} \\ & - Cas(e_{i,j})+ Bkg(x_i,x_j) \\
	with &\quad	\Delta x=|x_i-x_j|, \quad	\Delta y=|y_i-y_j|	
	\end{aligned}
\end{equation}
It includes regression factors ($a_0$, $a_1$, $a_2$, $a_3$, $b_0$, $b_1$) and considers the distance between the instances ($\Delta x$, $\Delta y$), the delay deduction for cascaded standard cells in one macro ($Cas(e_{i,j})$), and the extra routing delay for instances in different placement regions ($Bkg(x_i,x_j)$). The values of $Cas(e_{i,j})$ and $Bkg(x_i,x_j)$ are determined by look-up tables. To improve estimation accuracy, the timing estimation function is divided into several concatenated intervals based on the Euclidean distance between instances ($D_{Eucl}=\|(\Delta x , \Delta y)\|$). The regression factors may be different among these intervals, making $T_{net}(e_{i,j})$ a piece-wise function. This approach accounts for the fact of imbalanced dataset that the number of long-routing nets is significantly smaller than the number of short-routing nets in real designs for sampling. 

In the implementation of AMF-Placer~{2.0} for UltraScale FPGAs (xcvu095), the timing delay model ($T_{net}$) is divided into three intervals of Euclidean distance ($D_{Eucl}$): $[0,3)$, $[3,6)$, and $[6,+\infty)$. The regression factors $b_0$ and $b_1$ are found to be 0.3 and 0.5, respectively. The resulting regression model is visualized in Fig.\ref{timingDistribution}. The dataset shows a maximum delay of 3.469 ns and a minimum delay of 0 ns. The regression model based on these data has a mean error of 0.147 ns and a standard deviation of 0.225 ns. 	 Table\ref{cpdVeri} shows that the average relative error of our predicted critical path delay (CPD) is 8.59\%, compared to the Vivado post-route exact CPD. The average relative error of Vivado pre-route CPD estimation is 7.29\%. Our proposed net delay model is relatively optimistic because most of the samples in the dataset are not from congested regions, while critical paths usually route through congested regions.

In AMF-Placer 2.0, a light-weight parallel static timing analysis (STA) engine has been implemented based on the timing model for $T_{net}$ and $T_{logic}$, as well as the fundamental idea of OpenTimer~\cite{huang2015opentimer}. The netlist is treated as a directed acyclic graph with topological levelization, and the timing analysis of instances at the same level is performed in parallel. For each instance $v_i$, the actual arrival time will be:
\begin{equation}\label{arrivalTime}
	\begin{aligned}
		T_{arr}(v_i) = \max_{v_j \in fanin(v_i)}(T_{arr}(v_j)+T_{logic}(v_j)+T_{net}(e_{i,j}))
	\end{aligned}
\end{equation}
and the required arrival time will be:
\begin{equation}\label{reqTime}
	\begin{aligned}
		T_{req}(v_i) = \min_{v_j \in fanout(v_i)}(T_{req}(v_j)-T_{net}(e_{i,j})) - T_{logic}(v_i)
	\end{aligned}
\end{equation}
Accordingly, similar to~\cite{martin2019flat}~\cite{lin2020analytical}, for later timing optimization, the slack of a timing edge $e_t(i,j)$ between source instance $v_i$ and sink instance $v_j$ is defined as:
\begin{equation}\label{slack}
	\begin{aligned}
		Slack(e_t(i,j)) =& T_{req}(v_j)- T_{arr}(v_i)\\
		&- T_{logic}(v_i)-T_{net}(e_{i,j}) 
	\end{aligned}
\end{equation}
We can get a set of timing edges with negative timing slack $E_t = \left\{e_t(i,j)| 	Slack(e_t(i,j)) <0 \right\}$.

\begin{table}[t]
	\tiny
	\caption{Net Delay Model Validation with Benchmarks}
	\centering
	\def\arraystretch{1.2}
	\setlength\tabcolsep{1.6pt} 
	\begin{tabular}{|c|c|c|c|c|c|c|c|c|}
		\hline
		Benchmark Name                                                                 & BLSTM & \begin{tabular}[c]{@{}c@{}}Rosetta\\ DigitRecog\end{tabular} & \begin{tabular}[c]{@{}c@{}}Rosetta\\ FaceDetect\end{tabular} & SpooNN & MemN2N & Minimap2 & OpenPiton & Average \\ \hline
		AMF CPD Prediction (ns)                                                        & 7.23  & 9.62                                                         & 18.37                                                        & 9.34   & 9.62   & 7.61     & 11.23     &    -     \\
		$|$Relative Error$|$ (\%)                                                          & 15.56 & 8.94                                                         & 7.35                                                         & 6.39   & 9.83   & 4.43     & 7.64      & 8.59    \\ \hline
		\begin{tabular}[c]{@{}c@{}}Vivado Pre-Route\\ Prediction CPD (ns)\end{tabular} & 9.17  & 12.25                                                        & 18.59                                                        & 8.54  & 12.02 & 8.19    & 12.59    &    -     \\
		$|$Relative Error$|$ (\%)                                                          & 7.13  & 16.00                                                        & 6.25                                                         & 2.69   & 12.61  & 2.93     & 3.52      & 7.29    \\ \hline
		Actual Post-Route CPD (ns)                                                     & 8.56  & 10.56                                                        & 19.83                                                        & 8.78   & 10.67  & 7.96     & 12.16     &    -     \\ \hline
	\end{tabular}
	\label{cpdVeri}
	\setlength{\abovecaptionskip}{0cm}
	\setlength{\belowcaptionskip}{-0.5cm}
		\vspace{-0.5cm}
\end{table}
\subsection{Timing-driven Quadratic Placement}\label{timingQP}

AMF-Placer~{2.0}'s timing-driven placement strategy offers benefits in both local and global timing optimization. Locally, it adapts pseudo net weights based on timing slack values to resolve local timing violations. Globally, it uses a percentile-based evaluation of global timing quality and adjust the bias of the weights to improve timing at the design level.

\subsubsection{Problem Modeling Overview}
As mentioned in Section \ref{relatedWorks}, analytical placers approximate the wirelength (or HPWL) and some other metrics using numerical models for efficient solutions. Like many previous analytical placers\cite{anl0}\cite{UTPlaceF}\cite{ripplefpga}\cite{parallelpack}, to approximate the non-differentiable HPWL function, AMF-Placer uses weighted quadratic objective function $\widetilde{W_e}$, as follows:
\begin{equation}
	\widetilde{W_e} =\sum_{i,j \in e} [ w^{B2B}_{x,ij}(x_i-x_j)^2+w^{B2B}_{y,ij}(y_i-y_j)^2],\label{approximate}
\end{equation}
where $e$ is a net. Accordingly, $i$ and $j$ represent the instances connected to $e$.  Meanwhile $w^{B2B}_{x,ij}$ and $w^{B2B}_{y,ij}$ are weights set according to Bound2Bound net model\cite{b2b}. 
\begin{figure}[t]
	\setlength{\abovecaptionskip}{0cm}
	\setlength{\belowcaptionskip}{-0.5cm}
	\centering
	\includegraphics[width=0.8\linewidth]{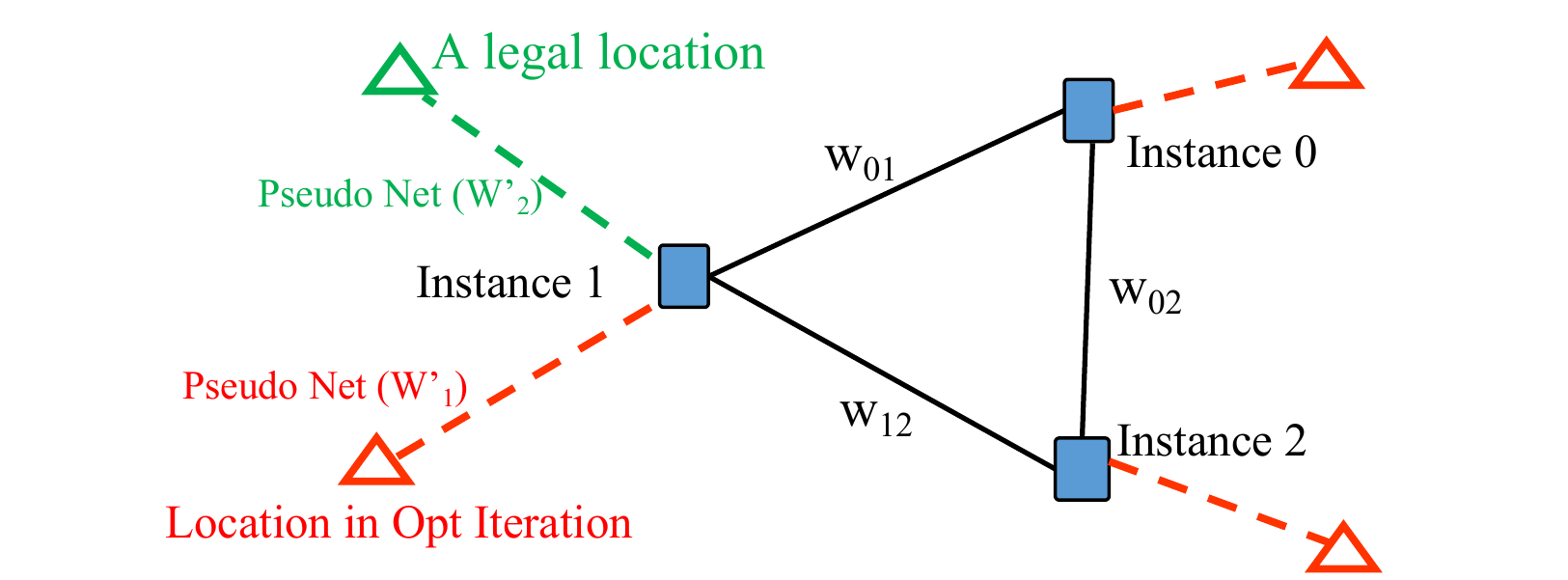}
	\caption{An Example of Pseudo Nets and Anchors: the dash lines indicate pseudo nets and the triangles represent anchors.} 
	\vspace{-0.5cm}
	\label{pseudoNet}
\end{figure}

To realize timing-driven placement, AMF-Placer 2.0 formulate a minimization problem as follows:
 \begin{equation}
	\begin{aligned}
		\min_{\mathbf{x},\mathbf{y}} \quad & (1-\lambda) W_{WL}(\mathbf{x},\mathbf{y})+  \lambda W_{T}(\mathbf{x},\mathbf{y}) \label{ourTimingOptProblem} \\
		\textrm{s.t.} \quad &  W_{WL}(\mathbf{x},\mathbf{y}) =   \sum_{e \in E} \widetilde{W_e} +  \sum_{e_p \in E_p} [w_{e_p} \widetilde{W_{e_p}}]\\
		 \quad &  W_{T}(\mathbf{x},\mathbf{y}) =    \sum_{e_b \in E_b} [w_{e_b}\widetilde{W_{e_b}}] +  \sum_{e_t \in E_t} [w_{e_t} \widetilde{W_{e_t}}] 
	\end{aligned}
\end{equation}
In this context, apart from the design net in $E$, there are three types of additional pseudo nets as shown in Fig.~\ref{pseudoNet}. 
\begin{itemize}
	\item $E_p$ represents the set of general pseudo nets used for legalization and density control~\cite{liang2021amf}, and $w_{e_p}$ denotes the specific weight assigned to each pseudo net.
	\item $E_b$ is a set of blockage-aware pseudo nets, as defined in Section~\ref{blockageSolution}, and its specific weight, $w_{e_b}$, is defined in Equation (\ref{guidenet}).
	\item $E_t$ is a set of timing-slack-aware pseudo nets, as defined in Section~\ref{timingModel}, and its weight, $w_{e_t}$, is related to the timing slack value of $e_t$, as specified in Equation (\ref{slack}).
\end{itemize}

$\widetilde{W_{e_b}}$, $\widetilde{W_{e_b}}$ and $\widetilde{W_{e_t}}$ are the HPWL quadratic approximation for the corresponding pseudo nets. Meanwhile, $\lambda$ balances the trade-off between wirelength and timing during the placement process, gradually transitioning from wirelength-driven placement to timing-driven placement. This allows the algorithm to achieve better timing performance while minimizing wirelength and area overheads.

\subsubsection{TNS/WNS-Aware Pseudo Nets} The pseudo nets in $E_t$ are pin-to-pin interconnections, and their HPWL approximation ($\widetilde{W_{e_t}}$) is simply the Manhattan distance between the two connected pins. This distance provides a basic estimate of the timing delay associated with the interconnection. The goal of minimizing this term $\sum_{e_t \in E_t} [w_{e_t} \widetilde{W_{e_t}}]$ is to reduce the approximate total negative slack (TNS), which is the sum of the negative slack of all the timing paths.  Since the timing edges in $E_t$ could be located in timing paths with different negative slacks, more weights should be assigned to $e_t$ in the most critical timing paths. In previous solutions~\cite{lin2021timing}~\cite{martin2019flat}, the criticality of $e_t$ was
\begin{equation*}
	w_{e_t}=\left(1-\frac{Slack(e_t)}{Dly_{max}}\right)^{\alpha} 
\end{equation*}
where $Dly_{max}$ represents the target delay value, and $\alpha$ is the constant criticality exponent that makes the weights adaptive to the nets with different criticalities. The distribution of timing slack can vary across different applications or stages of placement. For example, some control logic instances or macros with high fanout may have many tiny negative slacks and only a few significantly worse ones. This can make it difficult for the analytical model to improve the worst negative slack. Additionally, in most cases, only a small portion of paths violate timing constraints during placement iterations, making it unnecessary to set timing-aware pin-to-pin nets for all interconnections in the circuit.

Considering these factors, in contrast, in AMF-Placer~{2.0}, the criticality of $e_t$ is extended as follows:
\begin{equation}
	\label{ourCriticality}
	w_{e_t}=\left\{
	\begin{aligned}
		& 0  & (Slack(e_t) \geq 0) \\
		&\left(1-\frac{Slack(e_t)}{Dly_{max}}\right)^{C(Slack(e_t))} &   (Slack(e_t) < 0)
	\end{aligned}
	\right.
\end{equation}
with
\begin{equation}
	\label{globalCriticality}
	C(Slack(e_t))= max\left(\alpha,\frac{\beta Slack(e_t)}{T_{thr}}\right)
\end{equation}
where the parameters $\alpha$ and $\beta$ serve to restrict the possible values of the exponent, while $T_{thr}$ represents a percentile value used to evaluate timing quality globally. Specifically, $N_{thr}$ percent of edges exhibit negative timing slack worse than the threshold $T_{thr}$. This threshold is updated during STA before each quadratic placement iteration.  Based on Eqn(\ref{globalCriticality}) the weight $w_{e_t}$ will be set aggressively high for edges with significantly negative slack values exceeding $T_{thr}$. The self-adaptive exponent can address the WNS issue. For edges with less severe negative slack values better than $T_{thr}$, a relatively low weight value will be used. These small negative slack values will still be accumulated to minimize TNS. In practice, $\alpha$, $\beta$, and $N_{thr}$ are set to 0.9, 3, and 30, respectively.

Finally, the Eigen3 solver\cite{eigen} is adopted to handle the optimization for wirelength, TNS and WNS in Eqn.(\ref{ourTimingOptProblem}), with the high parallelism based on OpenMP.

\subsection{Global Packing}\label{globalPacking}

In the exact legalization after FPGA global placement, each instance must be mapped to a specific site on the FPGA, which contains a fixed number and type of resources. For example, LUTs, FFs, MUXes, and CARRYs with compatible signals can be grouped into a CLB site, reducing inter-site routing, as illustrated in Fig.~\ref{FPGA}.

A high-quality parallelized packing algorithm has been proposed by UTPlaceF\cite{parallelpack}, allowing FPGA sites to concurrently search for candidate packing solutions and negotiate during synchronization. AMF-Placer~{1.0} incorporated modifications to improve efficiency in this packing solution. In AMF-Placer~{2.0}, a timing factor was added to the priority evaluation of instances during parallel packing to optimize timing.

In the site-centric parallel packing algorithm, instances with locations near a site compete to occupy its resources. In the original implementation of \cite{parallelpack}, a score function is introduced as follows:
\begin{equation} 
	\begin{aligned}		
		\text {Score}_{UTP}\left({C, s_k}\right)=&\sum_{e \in \text {Net}\left ({C}\right)} \frac {\text {InternalPins}\left ({e, C}\right) - 1}{\text {TotalPins}\left ({e}\right) - 1} \\
		&-\,\,\theta \cdot \Delta \text {HPWL}\left ({C, s_k}\right) 		\label{utpPacking}
	\end{aligned}
\end{equation}
where
\begin{itemize} 
	\item  $C$ is a candidate cluster of instances
	\item  Net($C$) is the set of nets that have at least one cell in $C$
	\item  TotalPins($e$) is the total pin count of net $e$
	\item  InternalPins($e$,$C$) is the number of pins of net $e$ in $C$ \item  $\Delta \text {HPWL}\left ({C, s_k}\right)$ is the HPWL increase of moving the instances in $C$ from their global placement locations to site $s_k$
	\item  $\theta$ is a positive weighting parameter. 
\end{itemize} 

The priority for instance $v_i$ to be packed into the site $s_k$ is determined by the increase of $\text {Score}_{UTP}\left({C, s_k}\right)$ when inserting $v_i$  into $C$.

In Equation (\ref{utpPacking}), the first term aims to reduce inter-site routing, while the second term minimizes wirelength overhead. During this process, an instance $v_i$ may be mapped to a distant site due to resource contention. After timing-driven global placement, critical path instances should be packed into sites close to their original positions to preserve timing results. 

To achieve this, we utilize the longest length of paths including the instance as an extra metric during global packing. This is because placement variations in instances along a long path can accumulate and significantly degrade path delay, requiring significant effort during detailed placement to recover. For example, in benchmark FaceDetect~\cite{rosetta}, there are hundreds of timing paths with more than 50 instances. One path may have a delay of around 13ns after global placement, which may seem acceptable given the 15ns clock period constraint. However, if each instance in the path contributes a small delay increase of just 0.1ns, the path delay can dramatically increase to over 18ns, resulting in serious timing violations. 

Accordingly, we include an extra term when evaluating a candidate cluster as follows:
\begin{equation} 
	\begin{aligned}		
		\text {Score}_{AMF}=\text {Score}_{UTP}+\gamma \sum_{v_i \in C}\text{maxPathLen}(v_i)	\label{ourPackingP}
	\end{aligned}
\end{equation}
where \text{maxPathLen}($v_i$) is the longest length of paths including the instance and $\gamma$ is a parameter which we set to 0.05 empirically while $\theta$ in $\text {Score}_{UTP}$ is set to 0.01. 

Apart from the timing optimization, clock legalization is implemented as AMF-Placer~{1.0}. Additionally, there are multiple BEL slots in one CLB site as shown in Fig.~\ref{FPGA}. Therefore, for the instances mapped to the same CLB site, they will be mapped to the BEL slots by enumerating possible mapping solutions to minimize the maximum delay of involved timing paths.

\subsection{Detailed Placement}\label{detailedPlacement}

\begin{figure}[t]
	\vspace{-0.5cm}
	\setlength{\abovecaptionskip}{0cm}
	\setlength{\belowcaptionskip}{-0.5cm}
	\centering
	\includegraphics[width=\linewidth]{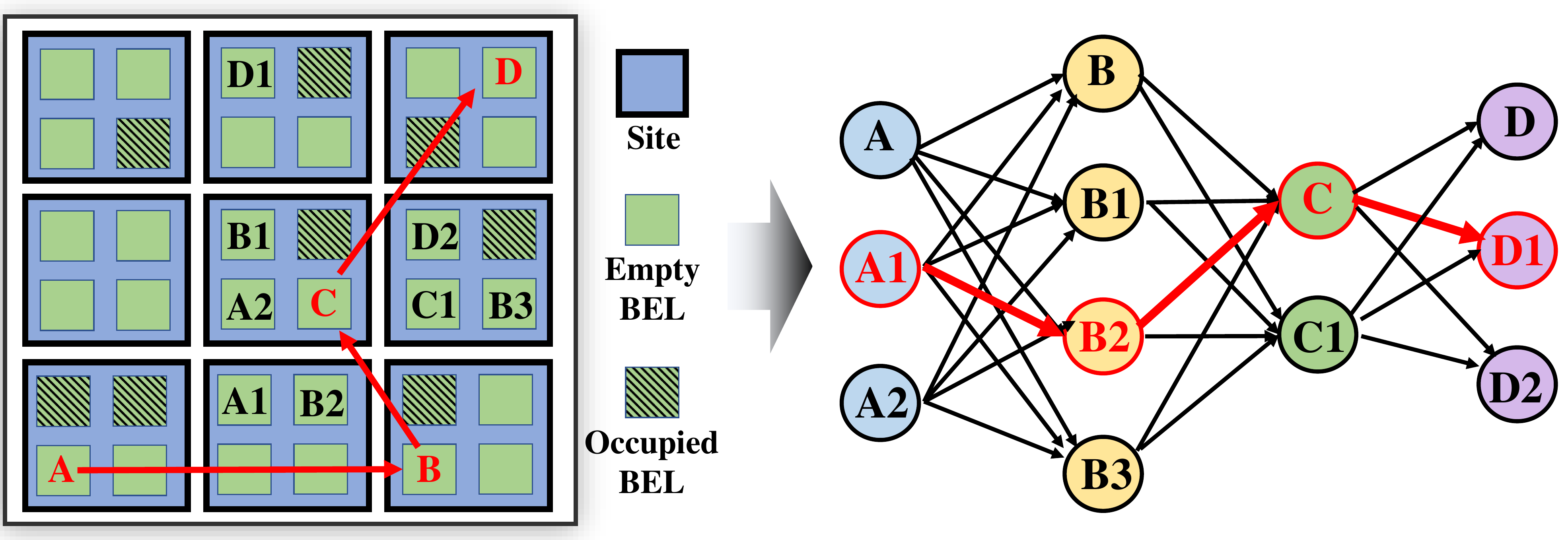}
	\caption{An example of shortest-path-based detailed placement: A, B, C and D have 3,4,2 and 3 candidate sites, respectively.} 
	
	\vspace{-0.5cm}
	\label{detailPE}
\end{figure}

\begin{figure}[t]
	\vspace{-0.5cm}
	\setlength{\abovecaptionskip}{-0.0cm}
	\setlength{\belowcaptionskip}{-0.5cm}
	\centering
	\includegraphics[width=\linewidth]{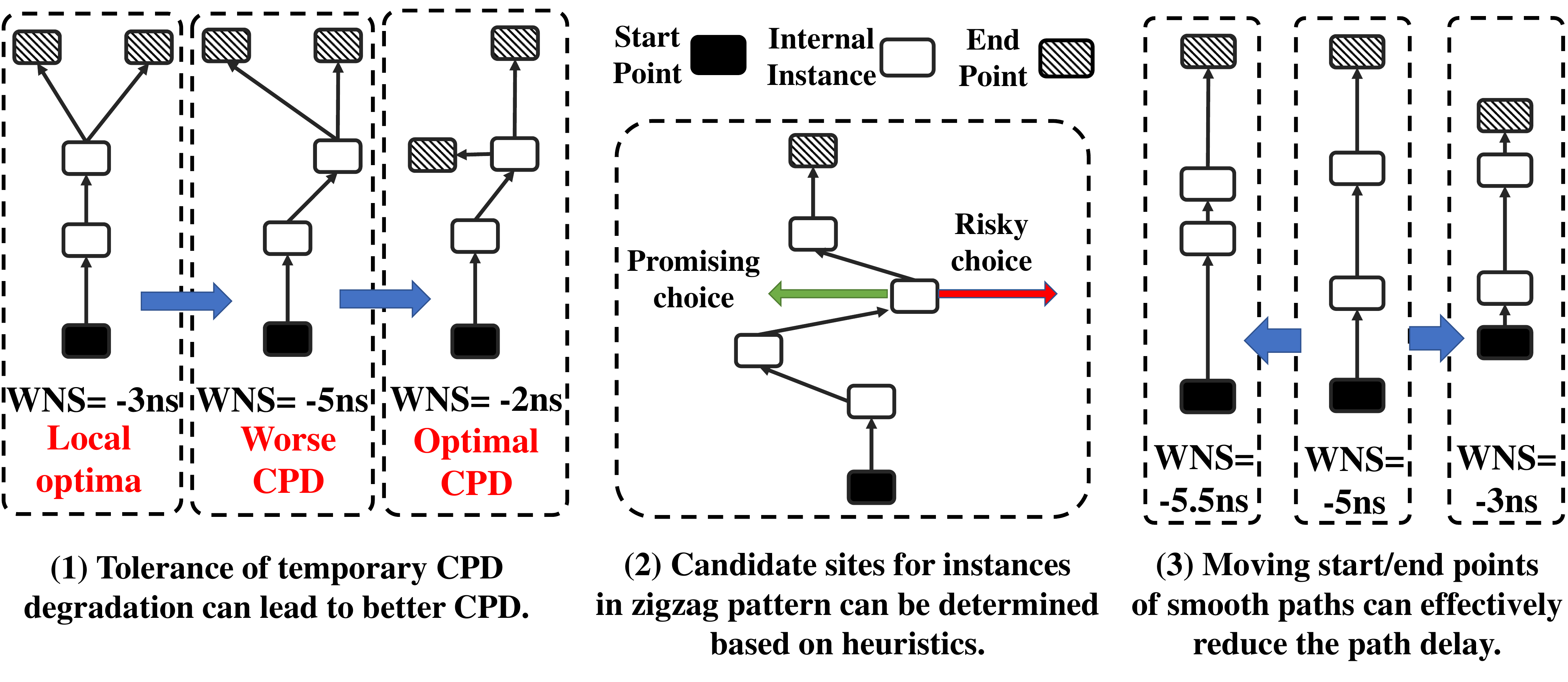}
	\caption{Examples explaining the motivation of detailed placement strategies}  
	\label{pathExample}
	\vspace{-0.5cm}
\end{figure}

Global placement and packing can hardly reach optimal timing delays of critical paths due to resource contention and mapping instances to specific sites. To solve this problem and minimize the worst negative slack (WNS), the detailed placement process is conducted locally on critical paths while minimizing interference with other instances. To do this, AMF-Placer~{2.0} adopts the widely-used shortest path algorithm\cite{dhar2017effective,martin2019flat,9570778}, which identifies the best candidate locations for each node on a timing path under optimization. This algorithm determines the location for each node to minimize the total delay of the target critical path. This process is illustrated in Fig.\ref{detailPE}, where a various number of candidate locations are identified for each node on the critical path A-$>$B-$>$C-$>$D. To solve this problem, a directed acyclic graph is constructed with multiple layers, each related to one instance in the critical path and containing vertices corresponding to candidate sites for the instance. Finding the shortest path from the source layer to the sink layer can then map the instances in the path to new sites and reduce the timing delay of this path. 
Compared to previous solutions, AMF-Placer 2.0 offers several key features. Firstly, it dynamically adjusts the scope of the target critical paths and the candidate sites for them, allowing for faster optimization with fewer iterations of incremental STA. Secondly, it tolerates a temporary increase in WNS during the shortest-path-based detailed placement, effectively resolving the problem related to local optima as shown in Fig.~\ref{pathExample}(1). Finally, it identifies promising candidates to reduce the number of candidate sites, reducing computational complexity and improving optimization.

\subsubsection{Self-Adaptation of Optimization Scope and Tolerance of Temporary Timing Degradation}

As shown in Algorithm~\ref{DPAlgo}, AMF-Placer~{2.0} conducts $N_{DPI}$ iterations of detailed placement, each processing $N_{CP}$ of the critical paths with the highest delay sequentially. For each target path, a conventional shortest path algorithm is applied. The instances in the previously optimized critical paths are marked fixed to preserve optimization results during the detailed placement for the less critical paths. $R_{nbr}$ controls the maximum distance between instances and their corresponding candidate sites, ranging from 1 to 0.1. At the beginning of each iteration, $R_{nbr}$ is decreased by $\Delta R$ to reduce the number of candidate sites for each instance, and $N_{CP}$ is increased by $\Delta N_{CP}$. By controlling $R_{nbr}$ and $N_{CP}$, more candidate sites and fewer target critical paths are used at the beginning of the detailed placement to realize fast improvement of CPD and WNS. In contrast, fewer candidate sites and more target critical paths are considered in the final iterations to fine-tune the locations of the instances and further improve timing optimization. The lowest CPD and its corresponding placement are recorded. If AMF-Placer 2.0 has not improved the CPD for $I_{thr}$ iterations, it reverts to the best recorded placement. This involves increasing $R_{nbr}$ by $(I_{thr}+1)\Delta R$ to provide more potential candidate sites for the most critical paths and reducing $N_{CP}$ by $(I_{thr}+1)\Delta N_{CP}$ to focus on the most critical paths. In an interval of $I_{thr}$ iterations, AMF-Placer~{2.0} tolerates the increase of CPD of the circuit. Empirically, $N_{DPI}$, $\Delta N_{CP}$, $\Delta R$, and $I_{thr}$ are set to 120, 20, 0.01, and 5, respectively.

	\begin{algorithm}[!b]
	\small
	\DontPrintSemicolon
	\KwIn{netlist information $H=(V,E)$, device information $DI$, instance locations after global packing $\mathbf{P}=\{(x^{pk}_i,y^{pk}_i)\}$}
	\KwOut{instance locations after optimization $\mathbf{P}'=\{(x^{dp}_i,y^{dp}_i)\}$}

	$\mathbf{P}'$ = $\mathbf{P}$
		
	bestCPD = staticTimingAnalysis($H$,$DI$,$\mathbf{P}'$);
	
	notOptCnt = 0; // recording the times of optimization failures
	
	$N_{CP}$, $R_{nrb}$ = 1, 1.0;

	\For{i=0; i$<N_{DPI}$; i++}
	{
		
		$L_{CP}$ = getMostCriticalPaths($N_{CP}$);
		
		\ForEach{path $\in$ $L_{CP}$} 
		{
			candMap = findCandidateLoc(path, $\mathbf{P}'$, $DI$, $R_{nrb}$);
			
			$\mathbf{P}'$ = shortestPath(path, candMap, $\mathbf{P}'$);
			
			markInstancesFixed(path);
		}
	
		markInstancesUnfixed($L_{CP}$);
		
		CPD = staticTimingAnalysis($H$,$DI$,$\mathbf{P}'$);
		
		$R_{nrb}$ -= $\Delta R$; $N_{CP}$ += $\Delta N_{CP}$; notOptCnt += 1;
		
		 \If{CPD$<$bestCPD}
		 {	
		 	bestCPD = CPD;
		 	
		 	recordTheBest($\mathbf{P}'$);
		 	
		 	notOptCnt = 0;
		 }
	 	\ElseIf{notOptCnt ==$ I_{thr}$} {
	 
			 recoverToBest($\mathbf{P}'$);
			 
			 notOptCnt = 0;
			 		
			 $R_{nrb}$ += $(I_{thr}+1)\Delta R$; 
			 
			 $N_{CP}$ -= $(I_{thr}+1)\Delta N_{CP}$;
 		}
	}

	recoverToBest($\mathbf{P}'$);
	
	optPerPathPerSingle\_DowngradeNotAllowed($H$,$DI$,$\mathbf{P}'$);
	
	\Return{$\mathbf{P}'$}
	
	\caption{Detailed Placement of AMF-Placer~{2.0}}

	\label{DPAlgo}
\end{algorithm}

\subsubsection{Smart Candidate Identification}
AMF-Placer 2.0 includes a sector-guided candidate selection algorithm for the concrete optimization of a specific path (i.e., line 8-9 in Algorithm~\ref{DPAlgo}). This algorithm identifies a smaller number of promising candidate sites for path delay optimization. Conventional solutions select candidate sites from a $d\times d$ square window around each instance on the path, which may be inefficient. AMF-Placer 2.0 makes two important observations based on the examples shown in Fig.~\ref{pathExample}: (1) for zigzag patterns in a path, beneficial candidate sites for an instance at a turning point  can be identified based on its predecessor and successor; (2) for a smooth or straight path, its overall delay is mainly determined by its start/end points.
\begin{figure}[t]
	\setlength{\abovecaptionskip}{-0.0cm}
	\setlength{\belowcaptionskip}{-0.5cm}
	\centering
	\includegraphics[width=\linewidth]{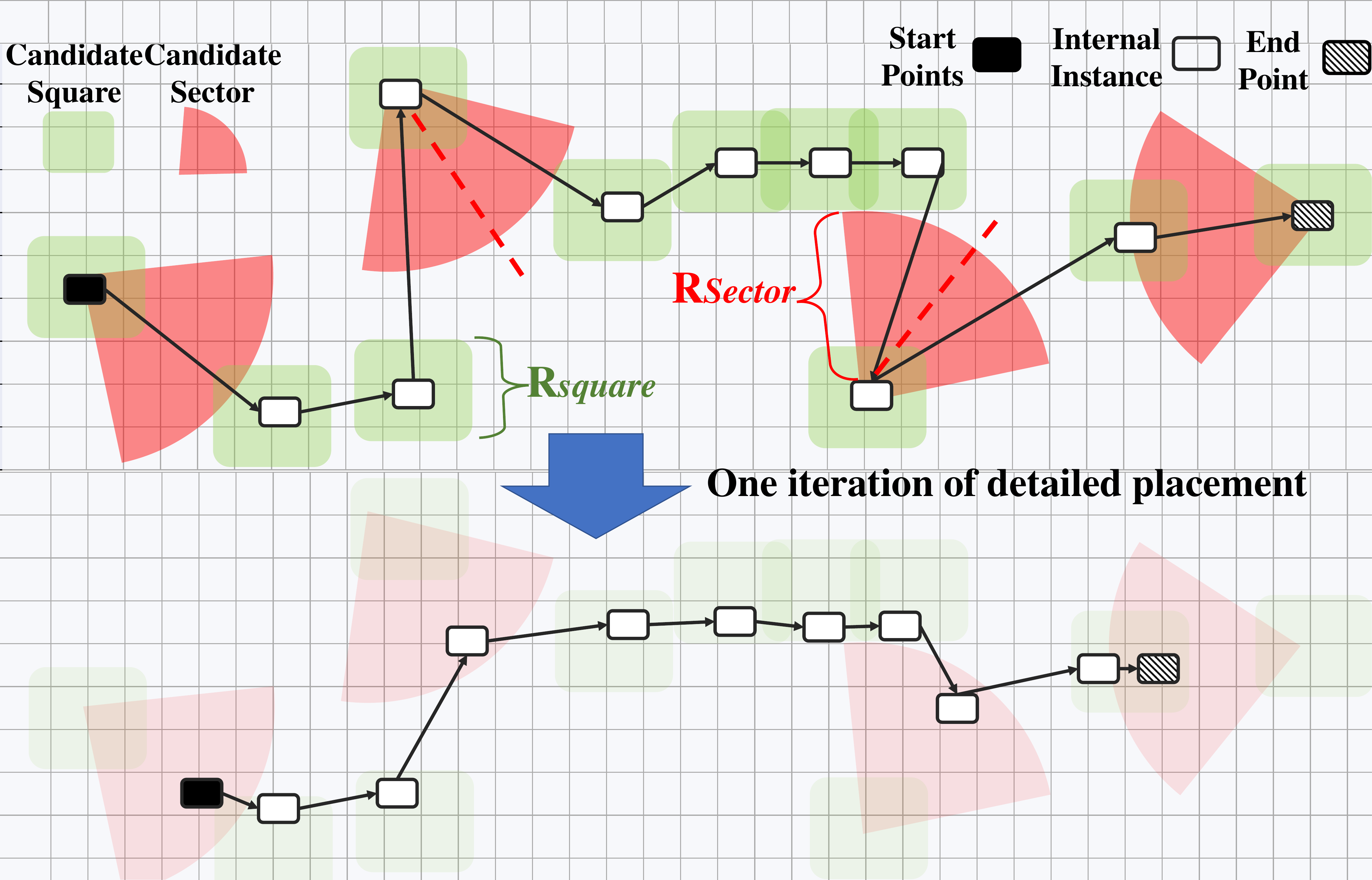}
	\caption{Examples of sector-based candidate site window: sector windows guide the instances move to promising directions.}  
	\label{sector}
	\vspace{-0.7cm}
\end{figure}

To identify candidates, AMF-Placer 2.0 uses small $d\times d$ square windows for all instances on the path, where $d=R_{nbr}R_{Square}$. Additionally, sector windows are set for specific instances, which could be start/end points or internal instances forming an acute triangle with their predecessor and successor. The center of a sector window overlaps with the corresponding instance and the radius is $R_{Sector}$, greater than $R_{Square}$. For internal instances, the direction of the window  aligns with the angle bisector of the triangle formed by the instance, its predecessor, and its successor, with the instance serving as the vertex. For start/end points, the direction of the sector window aligns with the direction vector from the instance to its successor/predecessor. An example is shown in Fig.~\ref{sector}. Overlapping candidate windows may lead to resource/architecture conflicts. If some BEL slots in these involved CLB sites conflict between instances, the assignment of BEL slots is resolved by balancing the candidate supply among instances, similar to \cite{dhar2017effective}. With this candidate selection algorithm, the runtime complexity can be significantly reduced. Empirically, $R_{Square}$ and $R_{Sector}$ are set to 3 and 5, respectively. In summary, the number of candidate sites for an instance is determined by both the global timing optimization progress and the local pattern of the instances on critical paths.

At the final stage of detailed placement, AMF-Placer~{2.0} will enable the constraint of WNS, i.e., no CPD increase will be accepted, to further fine-tune the placement. For detailed placement, an integrated parallel incremental timing analysis is implemented to fast evaluate the timing benefit.

Previous works\cite{dhar2017effective,martin2019flat,9570778} on placement optimization may suffer from limited timing optimization and high runtime overhead due to two reasons. Firstly, they are constrained by the requirement that no moves or swaps should increase the WNS, which may lead to local optima and limit the optimization. Secondly, the runtime complexity of the shortest path-based solutions is O($pL_{avg}N_{candidate}^2$)\cite{martin2019flat}, where $p$ is a small constant overhead factor, $L_{avg}$ is the average length of the critical paths considered, and $N_{candidate}$ is the average candidate site set size for critical path nodes. A small $N_{candidate}$ may limit the optimization space, while a large $N_{candidate}$ may significantly increase the runtime. In contrast, the solutions presented in AMF-Placer~{2.0} overcome these limitations effectively.

\section{Experimental Evaluation}\label{evaluation}

\subsection{Target Device, Benchmarks and Environment}\label{experimentSetting}

AMF-Placer~{2.0} targets AMD Xilinx VU095 FPGA devices and can transfer related techniques to other devices \cite{dhar2017effective}. The benchmarks used in the experiments are open-source, suitable for VU095 devices, and designed for various domains, such as CNN \cite{spoonn}, memory networks \cite{memn2n}, LSTM \cite{blstm}, SoC/NoC \cite{openpiton}, and genetic encode alignment \cite{minimap}. Benchmark OptimSoC \cite{optimsoc} used by AMF-Placer~{1.0} is removed since its major cause of timing violation is clock domain crossing, which is outside the scope of AMF-Placer~{2.0}. Table~\ref{benchmark} lists the parameters of the benchmarks and device, where macroRatio indicates the macro proportion of the design. Some of the benchmarks are generated via high-level synthesis \cite{rosetta}, while others are described in Verilog. Commercial IP cores in some benchmarks are black-box instances in the post-synthesis netlist and cannot be handled by RapidWright \cite{rapidwright} and other placers relying on the EDIF netlists. To handle general designs, AMF-Placer~{2.0} directly extracts the instance interconnection from Vivado via interactive Tcl commands.

AMF-Placer~{2.0} is implemented in C++ and experiments are conducted on Ubuntu 20.04 with an Intel i7-6770 CPU (3.40 GHz, 8 logic cores) and 32GB DDR4. All the experiments in this section use 8 threads.

\begin{table}[t]
	\tiny
	\caption{Parameters of Benchmarks and Device}
	\centering
	\def\arraystretch{1.2}
	\setlength\tabcolsep{2pt} 
	\begin{tabular}{|c|c|c|c|c|c|c|c|c|}
		\hline
		Benchmarks     & \begin{tabular}[c]{@{}c@{}}Rosetta\cite{rosetta} \\ FaceDetection\end{tabular} 
		& \begin{tabular}[c]{@{}c@{}}SpooNN\\ \cite{spoonn} \end{tabular} 
		& \begin{tabular}[c]{@{}c@{}}MiniMap2\\ \cite{minimap2} \end{tabular} 
		& \begin{tabular}[c]{@{}c@{}}OpenPiton\\ \cite{openpiton} \end{tabular}   
		& \begin{tabular}[c]{@{}c@{}}MemN2N\\ \cite{memn2n} \end{tabular}  
		& \begin{tabular}[c]{@{}c@{}}BLSTM\\ \cite{blstm} \end{tabular}    
		& \begin{tabular}[c]{@{}c@{}}Rosetta\cite{rosetta} \\ DigitRecog\end{tabular} 
		& \begin{tabular}[c]{@{}c@{}}Device:\\VU095\end{tabular} \\ \hline
		\#LUT          & 68945                                                           & 63095   & 407586   & 180388    & 184997 & 118967 & 151636    & 537600                                                           \\ \hline
		\#FF           & 56987                                                           & 70987   & 252624   & 111966    & 84694  & 54690  & 105580    & 537600                                                  \\ \hline
		\#CARRY        & 4978                                                            & 2091    & 19826    & 1712      & 11528  & 2762   & 1970      & 33600                                                        \\ \hline
		\#Mux          & 2177                                                            & 217     & 180      & 13696     & 4466   & 36210  & 4662      & 201600                                                         \\ \hline
		\#LUTRAM       & 255                                                             & 251     & 251      & 752       & 3500   & 1147   & 251       & 19200                                                       \\ \hline
		\#DSP          & 101                                                             & 165     & 528      & 58        & 312    & 258    & 1         & 768                                                       \\ \hline
		\#BRAM         & 141                                                             & 208        & 283      & 147       & 148    & 812    & 379    & 1728                                                         \\ \hline
		\#Cell         & 134450                                                          & 137937  & 681889   & 309145    & 289721 & 215101 & 265775    &  -                                                       \\ \hline
		\#Macro        & 3582                                                            & 1135       & 8746     & 8278      & 5775   & 14651  & 3061   &  -                                                          \\ \hline
		\#siteForMacro & 55666                                                           & 23079      & 191263   & 48066     & 118960 & 171822 & 55754   & -                                                          \\ \hline
		MacroRatio     & 40\%                                                            & 16\%       & 28\%     & 15\%      & 41\%   & 80\%   & 21\%    &-        	                        \\ \hline
		clockPeriod(ns)     & 15                          & 8       & 8     & 10     & 10  & 8   & 8   &-                                                       \\ \hline
	\end{tabular}
	\label{benchmark}
	\setlength{\abovecaptionskip}{0cm}
	\setlength{\belowcaptionskip}{-0.5cm}
		\vspace{-0.5cm}
\end{table}

\subsection{Comparison with Vivado}\label{comparisonWithVivado}

Existing open-source analytical FPGA placers do not support mixed-size FPGA placement of the aforementioned macros on Ultrascale devices. Therefore, for a comprehensive comparison, we used the widely-used commercial tool Xilinx Vivado 2020.2 and 2021.2 as our baselines. Notably, Xilinx Vivado 2021.2 employs machine learning, which is interesting for comparison.

The experimental results are presented in Table~\ref{vComparison}, where WNS represents the worst negative slack, CPD represents critical path delay, and RT represents the runtime of the entire placement flow. To evaluate the final timing quality, the placement results of AMF-Placer 2.0 were loaded by Vivado router to implement actual routing. Different combinations of placer and router may lead to different results, which is also evident in Table~\ref{vComparison}.

Here, AMF represents AMF-Placer 2.0, V2020 represents Vivado 2020.2, and V2021 represents Vivado 2021.2. For example, AMF-V2020 represents the combination of AMF-Placer 2.0 and the router of Vivado 2020.1. The WNS metrics, which are negative or near-zero in Table~\ref{vComparison}, indicate that the designer-defined timing constraints are strict for the placement. The CPD metrics mainly indicate the delay of the critical path. All the results of runtime and critical path delay are normalized to the results of Vivado 2020.2 for comparison, as indicated by the Rnorm values.
 \begin{table}[t]
	\centering
	\tiny
	\def\arraystretch{1.2}
	\setlength\tabcolsep{1.5pt} 
	\caption{Comparison of AMF-Placer~{2.0} with Vivado 2020.2 and 2021.2}
	\begin{tabular}{|c|ccccccccc|}
		\hline
		                            & \multicolumn{1}{c|}{Place-Route} &  \multicolumn{1}{c|}{BLSTM}  & \multicolumn{1}{c|}{DigitRecog} & \multicolumn{1}{c|}{FaceDetect} & \multicolumn{1}{c|}{SpooNN}  & \multicolumn{1}{c|}{MemN2N} & \multicolumn{1}{c|}{MiniMap2} & \multicolumn{1}{c|}{OpenPiton} & Average \\ \hline
		                            &            AMF-V2020             &            -0.389            &             -3.595              &             -0.384              &            -0.491            &           -1.601            &             0.049             &             -2.310             &    -    \\ \cline{2-10}
		          WNS(ns)           &            AMF-V2021             &            -0.508            &             -4.373              &             -0.445              &            -0.537            &           -1.665            &             0.001             &             -2.556             &    -    \\ \cline{2-10}
		                            &              V2020               &            -0.562            &             -2.564              &             -0.243              &            -0.779            &           -0.669            &             0.037             &             -2.159             &    -    \\ \cline{2-10}
		                            &              V2021               &            -0.668            &             -3.249              &             -0.264              &            -0.836            &           -0.732            &             0.070             &             -2.436             &    -    \\ \hline
		                            &      { \textbf{AMF-V2020}}       &             8.40             &              11.64              &              15.39              &             8.50             &            11.60            &             7.96              &             12.38              &    -    \\
		                            &       {\  \textbf{Rnorm}}        & {\color{red} \textbf{0.981}} &              1.097              &              1.009              & {\color{red} \textbf{0.967}} &            1.087            & {\color{red} \textbf{0.999}}  &             1.017              &  1.023  \\ \cline{2-10}
		                            &            AMF-V2021             &             8.51             &              12.41              &              15.45              &             8.55             &            11.66            &             8.01              &             12.56              &    -    \\
		          CPD(ns)           &              Rnorm               &            0.994             &              1.169              &              1.013              &            0.972             &            1.093            &             1.006             &             1.032              &  1.040  \\ \cline{2-10}
		                            &              V2020               &             8.56             &              10.61              &              15.24              &             8.79             &            10.67            &             7.96              &             12.17              &    -    \\
		                            &              Rnorm               &              1               &    {\color{red} \textbf{1}}     &    {\color{red} \textbf{1}}     &              1               &  {\color{red} \textbf{1}}   &               1               &    {\color{red} \textbf{1}}    &    1    \\ \cline{2-10}
		                            &              V2021               &             8.67             &              11.29              &              15.27              &             8.85             &            10.73            &             7.94              &             12.44              &    -    \\
		                            &              Rnorm               &            1.012             &              1.065              &              1.002              &            1.006             &            1.006            & {\color{red} \textbf{0.996}}  &             1.022              &  1.016  \\ \hline
		  \multicolumn{1}{|l|}{}    &               AMF                &             337              &               544               &               245               &             205              &             748             &              920              &              565               &    -    \\
		  \multicolumn{1}{|l|}{}    &              Rnorm               & {\color{red} \textbf{ 0.85}} &              1.04               &   {\color{red} \textbf{ 0.86}}  & {\color{red} \textbf{0.77}}  &            1.15             & {\color{red} \textbf{ 0.84}}  &              1.02              &  0.93   \\ \cline{2-10}
		\multicolumn{1}{|l|}{RT(s)} &              V2020               &             398              &               522               &               285               &             265              &             650             &             1094              &              555               &         \\
		  \multicolumn{1}{|l|}{}    &              Rnorm               &             1.00             &   {\color{red} \textbf{1.00}}   &              1.00               &             1.00             & {\color{red} \textbf{1.00}} &             1.00              &  {\color{red} \textbf{1.00}}   &  1.00   \\ \cline{2-10}
		  \multicolumn{1}{|l|}{}    &              V2021               &             396              &               529               &               300               &             271              &             759             &             1031              &              630               &         \\
		  \multicolumn{1}{|l|}{}    &              Rnorm               & 0.99                         &              1.01               &              1.05               &             1.02             &            1.17             &             0.94              &              1.14              &  1.05   \\ \hline
	\end{tabular}
	\vspace{-0.5cm}
	\label{vComparison}	
\end{table}

Based on the table, Vivado 2020.2 is currently the top-performing option in terms of both timing quality and runtime, and its router is particularly well-suited for use with AMF-Placer 2.0. However, it is worth noting that AMF-V2020 achieves only a 2.3\% higher CPD and 11.5\% lower runtime on average when compared to Vivado 2020.2, and a 0.69\% higher CPD and 7.0\% lower runtime on average when compared to Vivado 2021.2. It is important to keep in mind that the runtime comparison is only for reference, as AMF-Placer 2.0 skips certain post-placement optimizations such as LUT pin assignment and clock skew optimization. Nonetheless, AMF-Placer 2.0 is the first FPGA placer capable of handling timing-driven mixed-size placement of complex designs using various FPGA resources, and it achieves comparable quality to the latest commercial tools. According to the analysis of the concrete placement, we find that the major existing limitations of AMF-Placer~{2.0} can be categorized into to two types, listed as follows: 
\begin{itemize} 
	\item Timing estimation accuracy:As evaluated in Section~\ref{timingModel}, the timing estimation model used in AMF-Placer~{2.0} is relatively optimistic, as it does not consider some congested regions that overlap with critical paths. This issue is particularly noticeable for benchmarks DigitRecog~\cite{rosetta} and MemN2N~\cite{memn2n}, and may be addressed in the future by using machine-learning-based timing estimation approaches, such as \cite{9460398}. Additionally, the timing analysis at the floorplanning stage may be insufficient, leading to a poor initial placement. Moreover, a significant portion of the WNS for benchmark DigitRecog~\cite{rosetta} is caused by clock skew, which is not considered by existing works for the Ultrascale FPGA architecture. This problem may be resolved by using the approach proposed in \cite{zhu1994clock}.
	\item Design-aware factors:  The design netlists used in Vivado placement are hierarchical in nature. The design hierarchy information can be used to fully leverage regularity for the placement of local circuits, such as parallel accumulator datapaths, buses, and FIFOs. However, AMF-Placer~{2.0} uses a flattened netlist, which misses out on opportunities for regularity-related optimization. This limitation can be addressed by adopting approaches from existing works, such as \cite{dapa} and \cite{2dsoft}.
\end{itemize}

\subsection{Effectiveness of Proposed Optimization Techniques}

In this subsection, we evaluate the individual impact of the major proposed optimizations for different placement phases in AMF-Placer~{2.0}. Here we show their impact by setting the placer configurations as follows, to  disable different specified optimization technique:
\begin{itemize} 
	\item\emph{Cfg0}: all the optimization techniques are enabled as the configuration in Section~\ref{comparisonWithVivado}.
	\item\emph{Cfg1}: disable path-length-aware clustering before partitioning in Section~\ref{floorplanning}
	\item\emph{Cfg2}: disable blockage-aware spreading and anchor insertion in Section~\ref{blockageSolution}.
	\item\emph{Cfg3}: disable WNS-aware timing criticality pseudo net weight in Section~\ref{timingQP}, i.e.
	$C(Slack(e_t))=\alpha$
	\item\emph{Cfg4}: disable path-length-aware parallel packing in Section~\ref{globalPacking}
	\item\emph{Cfg5}: disable sector-guided site candidate selection in Section~\ref{detailedPlacement} and the value of $R_{Square}$ remains at 3 (i.e., just using the original small square window)
	\item\emph{Cfg6}: disable sector-guided site candidate selection in Section~\ref{detailedPlacement} and set $R_{Square}$ to be 5 (i.e., simply enlarge the square window)
\end{itemize}
\begin{table}[t]
	\centering
	\tiny
	\def\arraystretch{1.1}
	\setlength\tabcolsep{1.5pt} 
	\caption{Effectiveness of Proposed Optimization Techniques}
\begin{tabular}{|c|ccccccccc|}
	\hline
	                            & \multicolumn{1}{c|}{Configuration} &  \multicolumn{1}{c|}{BLSTM}   & \multicolumn{1}{c|}{DigitRecog} & \multicolumn{1}{c|}{FaceDetect} &  \multicolumn{1}{c|}{SpooNN}  &  \multicolumn{1}{c|}{MemN2N}  & \multicolumn{1}{c|}{MiniMap2} & \multicolumn{1}{c|}{OpenPiton} & Average \\ \hline
	                            &                Cfg0                &             8.40              &              11.64              &              15.39              &             8.50              &             11.60             &             7.96              &             12.38              &    -    \\
	                            &               Rnorm                & {\color{red} \textbf{ 1.000}} &  {\color{red} \textbf{ 1.000}}  &  {\color{red} \textbf{ 1.000}}  & {\color{red} \textbf{ 1.000}} & {\color{red} \textbf{ 1.000}} & {\color{red} \textbf{ 1.000}} &             1.000              &  1.000  \\ \cline{2-10}
	                            &                Cfg1                &             9.31              &              13.28              &              16.25              &             10.18             &             13.30             &             8.04              &             12.72              &    -    \\
	                            &               Rnorm                &             1.108             &              1.141              &              1.056              &             1.197             &             1.146             &             1.011             &             1.028              &  1.098  \\ \cline{2-10}
	                            &                Cfg2                &             8.58              &              12.36              &              18.00              &             10.07             &             13.42             &             8.55              &             13.52              &    -    \\
	                            &               Rnorm                &             1.021             &              1.062              &              1.170              &             1.185             &             1.157             &             1.075             &             1.092              &  1.109  \\ \cline{2-10}
	                            &                Cfg3                &             8.42              &              12.33              &              17.35              &             10.86             &             12.42             &             8.61              &             13.18              &    -    \\
	          CPD(ns)           &               Rnorm                &             1.003             &              1.060              &              1.127              &             1.278             &             1.071             &             1.082             &             1.065              &  1.098  \\ \cline{2-10}
	                            &                Cfg4                &             9.22              &              12.21              &              15.42              &             8.65              &             11.74             &             7.96              &             12.67              &    -    \\
	                            &               Rnorm                &             1.098             &              1.049              &              1.002              &             1.018             &             1.012             &             1.001             &             1.024              &  1.029  \\ \cline{2-10}
	                            &                Cfg5                &             8.48              &              11.78              &              15.45              &             8.50              &             11.80             &             7.97              &             12.35              &    -    \\
	                            &               Rnorm                &             1.010             &              1.012              &              1.004              &             1.000             &             1.017             &             1.001             & {\color{red} \textbf{ 0.998}}  &  1.006  \\ \cline{2-10}
	                            &                Cfg6                &             8.53              &              12.24              &              15.55              &             9.82              &             13.41             &             8.17              &             14.71              &    -    \\
	                            &               Rnorm                &             1.016             &              1.052              &              1.010              &             1.155             &             1.156             &             1.027             &             1.189              &  1.086  \\ \hline
	  \multicolumn{1}{|l|}{}    &                Cfg0                &              337              &               544               &               245               &              205              &              748              &              920              &              565               &    -    \\
	  \multicolumn{1}{|l|}{}    &               Rnorm                &             1.00              &              1.00               &              1.00               &             1.00              &             1.00              & {\color{red} \textbf{ 1.00}}  &              1.00              &  1.00   \\ \cline{2-10}
	  \multicolumn{1}{|l|}{}    &                Cfg1                &              326              &               528               &               323               &              189              &              743              &              986              &              550               &    -    \\
	  \multicolumn{1}{|l|}{}    &               Rnorm                & {\color{red} \textbf{ 0.97}}  &              0.97               &              1.32               & {\color{red} \textbf{ 0.92}}  &             0.99              &             1.07              &              0.97              &  1.03   \\ \cline{2-10}
	  \multicolumn{1}{|l|}{}    &                Cfg2                &              346              &               524               &               186               &              190              &              505              &              937              &              510               &    -    \\
	  \multicolumn{1}{|l|}{}    &               Rnorm                &             1.03              &              0.96               &  {\color{red} \textbf{ 0.76}}   &             0.93              &             0.68              &             1.02              &  {\color{red} \textbf{ 0.90}}  &  0.90   \\ \cline{2-10}
	  \multicolumn{1}{|l|}{}    &                Cfg3                &              351              &               483               &               279               &              196              &              445              &             1007              &              595               &    -    \\
	\multicolumn{1}{|l|}{RT(s)} &               Rnorm                &             1.04              &  {\color{red} \textbf{ 0.89}}   &              1.14               &             0.95              & {\color{red} \textbf{ 0.60}}  &             1.09              &              1.05              &  0.97   \\ \cline{2-10}
	  \multicolumn{1}{|l|}{}    &                Cfg4                &              355              &               527               &               245               &              204              &              729              &              966              &              558               &    -    \\
	  \multicolumn{1}{|l|}{}    &               Rnorm                &             1.05              &              0.97               &              1.00               &             1.00              &             0.97              &             1.05              &              0.99              &  1.00   \\ \cline{2-10}
	  \multicolumn{1}{|l|}{}    &                Cfg5                &              353              &               513               &               252               &              209              &              842              &              967              &              568               &    -    \\
	  \multicolumn{1}{|l|}{}    &               Rnorm                &             1.05              &              0.94               &              1.03               &             1.02              &             1.12              &             1.05              &              1.00              &  1.03   \\ \cline{2-10}
	  \multicolumn{1}{|l|}{}    &                Cfg6                &              381              &               609               &               246               &              195              &              826              &              998              &              656               &    -    \\
	  \multicolumn{1}{|l|}{}    &               Rnorm                &             1.13              &              1.12               &              1.01               &             0.95              &             1.10              &             1.09              &              1.16              &  1.08   \\ \hline
\end{tabular}

\vspace{-0.5cm}
	\label{individualEvaluation}	
\end{table}

As shown in Table~\ref{vComparison}, it is evident that AMF-Placer~{2.0} performs better when paired with the router of Vivado 2020.2. Therefore, we have chosen the "AMF-V2020" configuration as baseline for the comparisons.

Table~\ref{individualEvaluation} presents the results of the experiments conducted with different configurations, with all results normalized to those of \emph{Cfg0} for ease of comparison, as indicated by the Rnorm values. 
For most situations, enabling all the proposed optimization techniques (\emph{Cfg0})  can realize better timing quality.

\subsubsection{Timing Quality of Global Placement} When optimization techniques,  such as \emph{Cfg1}, \emph{Cfg2}, and \emph{Cfg3}, are disabled, the timing quality can degrade by up to 10\%. These results indicate that detailed placement optimization cannot compensate for low-quality global placement. Specifically, disabling blockage-aware spreading and anchor insertion (\emph{Cfg2}) leads to an average increase of 10.9\% in CPDs for benchmarks, underscoring the significant impact of placement blockages and the importance of corresponding optimizations. 

\subsubsection{Runtime of Global Placement} Disabling optimization techniques for global placement may reduce runtime by conducting fewer analyses and processes. For instance, disabling blockage-aware spreading and anchor insertion results in an average 10\% reduction in runtime for benchmarks since the involved DFS-based clustering is not parallelized. However, despite the resulting runtime overheads, these optimization techniques are justified given the benefits they offer.

\subsubsection{Timing Quality of Packing and Detailed Placement} Disabling the corresponding techniques, such as \emph{Cfg4}, \emph{Cfg5}, and \emph{Cfg6}, can significantly degrade both the timing quality and runtime of placement. Results obtained with \emph{Cfg4} show that removing the path-length-based factor from Eqn.(\ref{ourPackingP}) for global packing increases CPDs by an average of 2.9\%. This is because critical paths containing many instances are noticeably disturbed during final packing. Detailed placement using small candidate windows, as implemented in \emph{Cfg5}, results in longer runtimes and fails to reach the optimal critical path delay for most benchmarks. In contrast, sector-guided candidate windows can achieve promising location updates while preserving the proper locations of other instances. On the other hand, using large square candidate windows in \emph{Cfg6} can degrade the timing quality. This is because the serious overlaps of the large windows lead to many candidate site conflicts between instances in one critical path \cite{dhar2017effective}. 

\begin{figure}[t]
	\setlength{\abovecaptionskip}{-0.0cm}
	\setlength{\belowcaptionskip}{-0.5cm}

	\centering
	\includegraphics[width=0.8\linewidth]{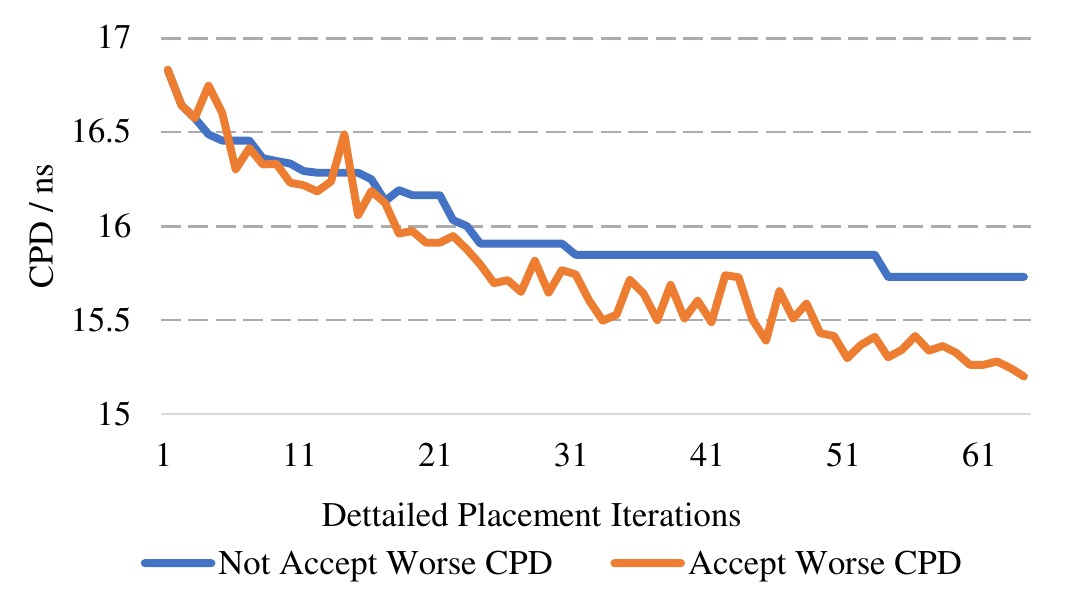}
	\caption{Example of benchmark FaceDetect~\cite{rosetta} shows the benefits of tolerating the temporary degradation of timing during detailed placement.}  
	\label{CPDTrace}
		\vspace{-0.5cm}
\end{figure}

In Section~\ref{detailedPlacement}, we discussed the importance of temporarily accepting worse CPD during detailed placement to escape from local optima and reach the optimal CPD of final placement. Fig.\ref{CPDTrace} shows two traces of CPD of benchmark FaceDetect\cite{rosetta} during detailed placement. One of them is obtained without allowing CPD degradation during detailed placement, while the other is obtained by temporarily accepting timing downgrading. Although the strict constraint of CPD descending can lead to a smooth CPD optimization trace, tolerating temporary degradation can result in a much better final result.

\subsubsection{Runtime of Packing and Detailed Placement} Removing the path-length-based factor from Eqn.(\ref{ourPackingP}) for global packing, as implemented in \emph{Cfg4}, can significantly increase runtime. This is because there are many zigzag critical paths that require a lot of runtime for detailed placement to handle them. Using small square windows without sector windows, as implemented in \emph{Cfg5}, can also increase runtime since these narrow windows slow down the optimization progress of critical paths. Meanwhile, the results of timing quality with large square windows for candidate selection (\emph{Cfg6}) can be realized by using sector windows for candidate selection with lower runtime.

\subsection{Portability to Commercial Tools}

A set of toolchains is available for interacting with Vivado, such as extracting information on various netlist and device models, which can assist users in handling other designs and devices.

Placement generated by AMF-Placer can be loaded into Vivado via a Tcl script for routing, and all placements generated by the proposed placement flow in this paper can be successfully routed. However, due to license restrictions in the Vivado patch used in the ISPD 2015/2016 contests, we are unable to obtain the routed wirelength from Vivado despite the availability of timing reports.

For more detailed statistics and usage of AMF-Placer 2.0, please refer to the open-source project documentation where we hope that the tool can help people in the community keep up with the progress of commercial tools.

\section{Conclusion}
In this paper, we introduce AMF-Placer 2.0, an open-source FPGA placer for complex practical designs with various FPGA resources. AMF-Placer 2.0 utilizes new techniques for timing optimization, including a simple timing model, placement-blockage-aware anchor insertion, WNS-aware timing-driven quadratic placement, and sector-guided detailed placement. Experimental results show that AMF-Placer 2.0 achieves only a 2.3\% higher critical path delay (CPD) and 11.5\% lower runtime on average than Vivado 2020.2, and a 0.69\% higher CPD and 7.0\% lower runtime on average than Vivado 2021.2.  The source code, Wiki, open-source benchmarks, and checkpoint/log files for the experiments are available at \href{https://github.com/zslwyuan/AMF-Placer}{https://github.com/zslwyuan/AMF-Placer}.

\section*{Acknowledgment}
The authors sincerely appreciate the kindly suggestions from reviewers, detailed explanations of UTPlaceF\cite{parallelpack} 
from Dr. Wuxi Li, and useful advice on Vivado metric usages from Dr. Stephen Yang\cite{ispd2016}, 
and detailed feedbacks from Mr. Jing Mai, the gold user of AMF-Placer~{1.0}. Furthermore, the authors acknowledge support from the Hong Kong Research Grants Council General Research Fund (GRF No. 16213521).


\ifCLASSOPTIONcaptionsoff
  \newpage
\fi



%
\bibliography{sample-bibliography}

%
%

\begin{IEEEbiography}[
{
\includegraphics[width=1.4in,height=1.25in,clip,keepaspectratio]{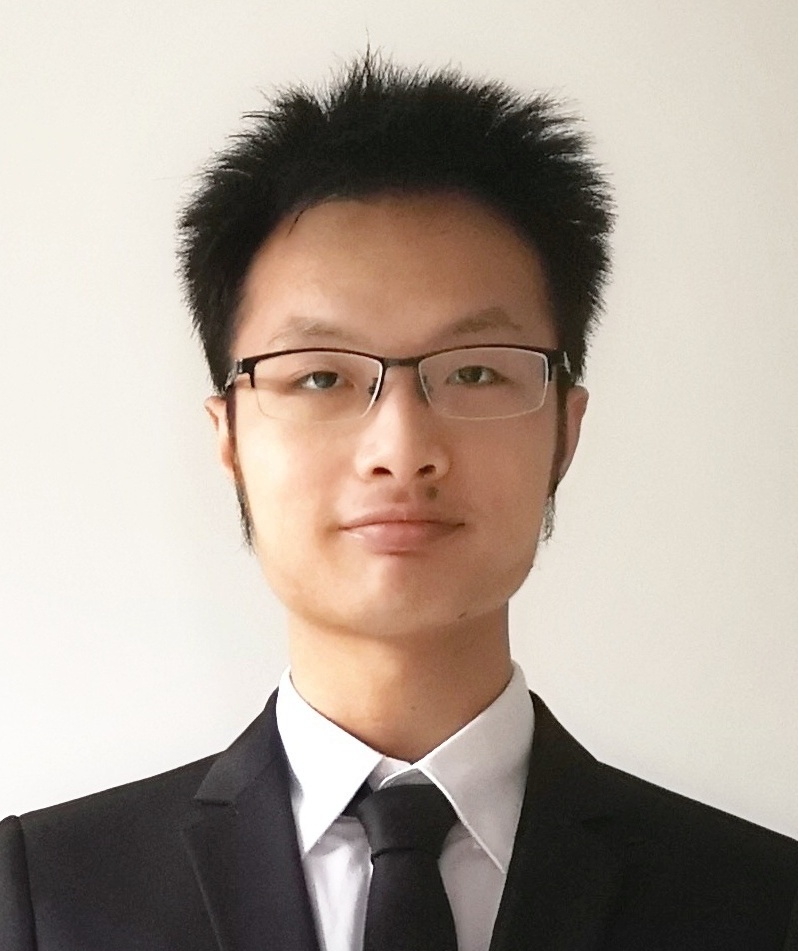}
}
]{Tingyuan Liang}
 received his B.S. degree in Electrical and Information Engineering from Zhejiang University in 2017. After completing his undergraduate studies, he pursued advanced degrees and earned his MPhil and PhD degrees in the Department of Electronic and Computer Engineering at the Hong Kong University of Science and Technology in 2019 and 2023, respectively.  His current research interests include EDA algorithms and computer architecture.

\end{IEEEbiography}
\vskip 0pt plus -1fil
\vspace*{-2\baselineskip}
\begin{IEEEbiography}[
	{
		\includegraphics[width=1in,height=1.25in,clip,keepaspectratio]{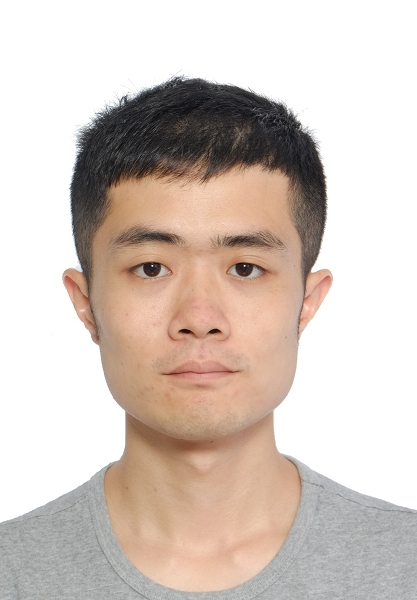}
	}
	]{Gengjie Chen} obtained his Ph.D. degree, under the supervision of Prof. Evangeline F. Y. Young, from the Department of Computer Science and Engineering of The Chinese University of Hong Kong (CUHK) in 2019. Before that, he received his bachelor degree in the Department of Electronic and Communication Engineering from Sun Yat-sen University (SYSU) in 2015. His research interests include electronic design automation (EDA), combinatorial optimization, numerical optimization, and machine learning.
\end{IEEEbiography}
\vskip 0pt plus -1fil
\vspace*{-2\baselineskip}
\begin{IEEEbiography}[
{
\includegraphics[width=1in,height=1.25in,clip,keepaspectratio]{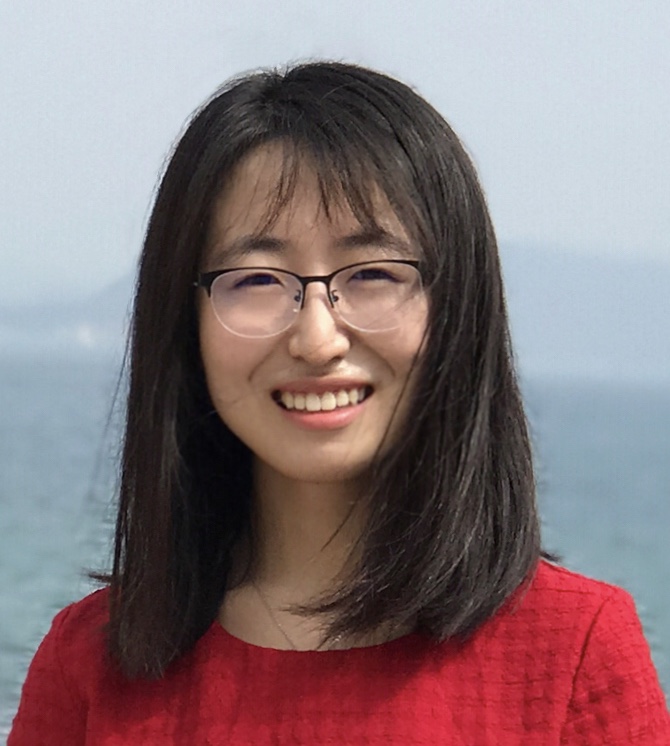}
}
]{Jieru Zhao}

(M’22) received the Ph.D. degree in
electronic and computer engineering from the Hong
Kong University of Science and Technology, Hong
Kong, in 2020. She is an Assistant Professor with the
Department of Computer Science and Engineering,
Shanghai Jiao Tong University, Shanghai, China.
Her current research interests include reconfigurable
system, high-level synthesis and hardware-software
co-design for emerging applications. Dr. Zhao was
a recipient of the Best Paper Award at ICCAD 2017
and Best Paper Nominations at DATE 2022, SC
2021, and CASES 2018.

\end{IEEEbiography}
\vskip 0pt plus -1fil
\vspace*{-2\baselineskip}
\begin{IEEEbiography}[
{
\includegraphics[width=1in,height=1.25in,clip,keepaspectratio]{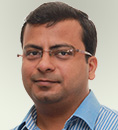}
}
]{Sharad Sinha}
(S’03, M’14) is an associate professor
with Dept. of Computer Science and Engineering,
Indian Institute of Technology (IIT) Goa. Previously,
he was a Research Scientist at NTU, Singapore. He
received his PhD degree in Computer Engineering
from NTU, Singapore (2014). He received the Best
Speaker Award from IEEE CASS Society, Singa-
pore Chapter, in 2013 for his PhD work on High
Level Synthesis and serves as an Associate Editor
for IEEE Potentials and ACM Ubiquity. Dr. Sinha
earned a Bachelor of Technology (B.Tech) degree
in Electronics and Communication Engineering from Cochin University of
Science and Technology (CUSAT), India in 2007. From 2007-2009, he was a
design engineer with Processor Systems (India) Pvt. Ltd. Dr. Sinha’s research
and teaching interests are in computer architecture, embedded systems and
reconfigurable computing.
\end{IEEEbiography}
\vskip 0pt plus -1fil
\vspace*{-2\baselineskip}
\begin{IEEEbiography}[
{
\includegraphics[width=1in,height=1.25in,clip,keepaspectratio]{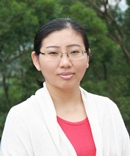}
}
]{Wei Zhang}
Wei Zhang (M’05) received a Ph.D. degree from
Princeton University, Princeton, NJ, USA, in 2009.
She was an assistant professor with the School
of Computer Engineering, Nanyang Technological
University, Singapore, from 2010 to 2013. Dr. Zhang
joined the Hong Kong University of Science and
Technology, Hong Kong, in 2013, where she is
currently a professor and she established the reconfigurable computing system laboratory (RCSL).
Dr. Zhang has authored or co-authored over 80 book chapters and papers in peer reviewed journals and international conferences. Dr. Zhang serves as the Associate Editor for TECS, TVLSI, and JETC.
She also serves on many organization committees and technical program
committees. Dr. Zhang’s current research interests include reconfigurable sys-
tems, FPGA-based design, low-power high-performance multicore systems,
electronic design automation, embedded systems, and emerging technologies.
\end{IEEEbiography}

\end{document}